\documentclass[aps,prl,twocolumn,superscriptaddress,showpacs]{revtex4-2}
\usepackage[latin1,utf8]{inputenc}
\usepackage[pdftex]{epsfig}
\usepackage{graphicx}
\usepackage{color}
\usepackage{latexsym}
\usepackage{amssymb}
\usepackage{amsmath}
\usepackage{longtable}
\usepackage{IEEEtrantools}
\usepackage{subfigure}
\usepackage{blkarray}
\usepackage{mathtools}
\usepackage[pdftex,colorlinks=true,allcolors=blue]{hyperref}
\usepackage[dvipsnames]{xcolor}
\usepackage[normalem]{ulem}
\begin{document}
 
\title{Long-range magnetic interactions in Nd$_2$PdSi$_3$ and the formation of skyrmion phases in centrosymmetric metals}

\author{Viviane~Pe\c{c}anha-Antonio}
\email{viviane.antonio@stfc.ac.uk}
\affiliation{Department of Physics, University of Oxford, Clarendon Laboratory, Oxford OX1 3PU, United Kingdom}
\affiliation{ISIS Neutron and Muon Source, STFC Rutherford Appleton Laboratory, Didcot OX11 0QX, United Kingdom}
\author{Zhaoyang Shan}
\author{Michael~Smidman}
\affiliation{Center for Correlated Matter and School of Physics, Zhejiang University, Hangzhou 310058, China}
\author{Juba Bouaziz}
\affiliation{Department of Physics, The University of Tokyo, Bunkyo-ku, Tokyo, 113-0033, Japan}
\author{Bachir~Ouladdiaf}
\author{Iurii~Kibalin}
\author{Marie-H\'{e}l\`{e}ne~Lem\'{e}e-Cailleau}
\affiliation{Institut Laue-Langevin, 6 Rue Jules Horowitz, BP 156, 38042 Grenoble Cedx 9, France}
\author{Christian~Balz}
\affiliation{ISIS Neutron and Muon Source, STFC Rutherford Appleton Laboratory, Didcot OX11 0QX, United Kingdom}
\author{Jakob~Lass}
\affiliation{PSI Center for Neutron and Muon Sciences, 5232 Villigen PSI, Switzerland}
\author{Daniel~A.~Mayoh}
\author{Geetha~Balakrishnan}
\author{Julie B. Staunton}
\affiliation{Department of Physics, University of Warwick, Coventry CV4 7AL, United Kingdom}
\author{Devashibhai~Adroja}
\affiliation{ISIS Neutron and Muon Source, STFC Rutherford Appleton Laboratory, Didcot OX11 0QX, United Kingdom}
\affiliation{Highly Correlated Matter Research Group, Physics Department, University of Johannesburg, Auckland Park 2006, South Africa}
\author{Andrew~T.~Boothroyd}
\email{andrew.boothroyd@physics.ox.ac.uk}
\affiliation{Department of Physics, University of Oxford, Clarendon Laboratory, Oxford OX1 3PU, United Kingdom}

\date{\today}

\begin{abstract}
We present an extensive X-ray and neutron scattering study of the structure and magnetic excitations of Nd$_2$PdSi$_3$, a sister compound of Gd$_2$PdSi$_3$ which was recently found to host a skyrmion lattice phase despite its centrosymmetric crystal structure. Dispersive magnetic excitations were measured throughout the Brillouin zone and modeled to determine the magnetic interactions between Nd ions. 
Our analysis reveals that the magnetic interactions in this system extend over large distances and are significantly affected by a crystallographic superstructure formed by ordering of the Pd and Si atoms. The results suggest that the mechanism for the skyrmion phase formation in this family of materials, specifically Gd$_2$PdSi$_3$, is through the long-range RKKY interactions rather than short-range triangular-lattice frustration.
\end{abstract}

\maketitle

The discovery of magnetic skyrmions in non-centrosymmetric compounds with the B20 crystal structure \cite{Muhlbauer,PhysRevLett.102.197202,Yu} has revolutionised the study of spin topology in condensed matter. More than that, it has initiated a new field of research, \textit{skyrmionics}, which aims to harness the exotic properties resulting from the coupling between magnetic and electronic degrees of freedom in these materials \cite{Fert2013,Lancaster2019,Bogdanov2020}. In proposals for a new generation of memory and logic devices based on skyrmion crystals (SkX), small skyrmion sizes, of the order of a few unit cells, are preferable over spin textures extending over large crystallographic scales \cite{Thomas,Wiesendanger,Moreau-Luchaire}. Recent conceptual qubit constructions, for example, rely upon the information density a nanoscale SkX would be able to store \cite{PhysRevLett.127.067201,PhysRevLett.130.106701}.

Theoretical work has helped broaden the spectrum of candidate skyrmion crystals, to the extent that predictions for topologically non-trivial magnetism exist even for centrosymmetric compounds
%to the extent that predictions exist for topologically non-trivial magnetism even in centrosymmetric compounds
\cite{PhysRevLett.108.017206,Leonov2015,PhysRevX.9.041063,PhysRevLett.124.207201,PhysRevB.103.104408,PhysRevB.95.224424,PhysRevLett.125.117204,PhysRevB.93.064430,PhysRevB.105.054435,PhysRevB.103.024439}. This is perhaps surprising given that the Dzyaloshinskii--Moriya interaction (DMI) usually responsible for stabilizing such phases is absent in crystals with inversion symmetry. In place of DMI a variety of different mechanisms have been proposed, some of which combine superexchange interactions with geometrical frustration \cite{PhysRevLett.108.017206,Leonov2015,PhysRevX.9.041063,PhysRevB.93.064430} while others show that topological spin textures can be stabilised even in the absence of frustration or magnetic field \cite{PhysRevB.103.104408,PhysRevB.95.224424,PhysRevB.103.024439}. Most theories involve exchange couplings of a long-range nature, requiring two or three nearest neighbours \cite{PhysRevLett.108.017206,Leonov2015,PhysRevX.9.041063,PhysRevLett.124.207201,PhysRevB.103.104408,PhysRevB.95.224424,PhysRevB.105.054435,PhysRevB.103.024439}. For metallic systems, the Ruderman-Kittel-Kasuya-Yosida (RKKY) type coupling provides a natural mechanism for the formation of skyrmion-like multi-\textbf{q} structures via Fermi-surface nesting \cite{PhysRevLett.124.207201,PhysRevB.103.104408,PhysRevB.104.184432,PhysRevLett.128.157206,PhysRevLett.133.016401,PhysRevB.103.024439}. Importantly, skyrmions in centrosymmetric materials are predicted to be much smaller and more densely packed than in non-centrosymmetric crystals.

Experiments have now confirmed the existence of a handful of such candidate centrosymmetric skyrmion compounds \cite{Kurumaji_science,Hirschberger_2019,Khanh_2020,Takagi2022,Gao_2020}. One of these is Gd$_2$PdSi$_3$, a member of a wider family of $R_{2}$PdSi$_{3}$ intermetallics with stacked triangular layers of $R$ (rare-earth) atoms which have been extensively studied for their exotic magnetic properties \cite{Mallik_1998,KOTSANIDIS1990199,MALLIK1998169,PhysRevB.62.425,SZYTULA1999365,Frontzek_2007,FRONTZEK2006398,PhysRevB.68.012413,PhysRevB.62.14207,PhysRevB.64.012418,PhysRevB.67.212401,PhysRevB.60.12162}. Besides its anisotropic magnetic behaviour and large negative magnetoresistance \cite{PhysRevB.60.12162,PhysRevLett.134.046702}, Gd$_2$PdSi$_3$ was shown to host a skyrmion lattice phase in moderate applied magnetic fields \cite{Kurumaji_science}. The SkX was proposed to be a triple-\textbf{q} spin structure formed by the superposition of three spin helices at 120$^\circ$ to each other \cite{Kurumaji_science}. 

To understand the formation of a SkX in Gd$_2$PdSi$_3$ and other centrosymmetric metals, it is essential to quantify the magnetic interactions. This can be achieved via inelastic neutron scattering (INS) measurements of dispersive magnetic excitations, but for Gd$_2$PdSi$_3$ this is not straightforward. First, the magnetic order is complex. Second, the Pd and Si atoms form a superstructure which creates many inequivalent exchange paths. This superstructure lowers the hexagonal space group symmetry and has a significant influence on the electronic bands near the Fermi level \cite{PhysRevLett.133.016401}, implying that any conduction-electron-mediated exchange must also be affected. Third, samples containing Gd are extremely challenging to study by INS owing to the large neutron absorption cross-section of Gd. The only neutron scattering studies reported so far used samples enriched with $^{160}$Gd to reduce neutron absorption \cite{PhysRevLett.129.137202,Ju2023}. In Ref.~\onlinecite{PhysRevLett.129.137202}, the authors analysed INS and paramagnetic diffuse scattering data from a polycrystalline sample to develop a minimal model, and found evidence that the magnetic interactions extend beyond the first few nearest neighbors.

In this work we take a different approach, choosing instead to study the related non-SkX compound Nd$_2$PdSi$_3$, whose ferromagnetic structure is amenable to analysis, and whose neutron absorption is sufficiently small that high-quality INS data could be obtained from a single-crystal sample. We determine the Pd/Si superstructure in Nd$_2$PdSi$_3$ and take it into account when modelling the magnetic spectrum. We find clear evidence that the interactions extend over long distances, switching from mainly ferromagnetic to antiferromagnetic coupling with increasing atomic separation.  Assuming the magnetic interactions are qualitatively similar in the Nd and Gd compounds (see below), our results support the theory that the SkX in centrosymmetric Gd$_2$PdSi$_3$ is most probably stabilised by long-range RKKY interactions.

When first reported, the $R_{2}$PdSi$_{3}$ intermetallics were believed to adopt an AlB$_2$-type hexagonal structure with space group $P6/mmm$ \cite{SZYTULA1999365}. In this description, $1a$ is the Wyckoff position of the $R$ atoms, while the Pd/Si atoms randomly occupy the $2d$ sites. More recent single crystal work demonstrated that Pd and Si in the heavy rare-earth members are in fact arranged in a $ 2a \times 2a \times 8c$ superstructure, where $a$ and $c$ are the hexagonal lattice parameters (see \cite{PhysRevB.84.104105} and references therein). In the present work \cite{suppl}, we find that the supercell in Nd$_{2}$PdSi$_{3}$ has dimensions $2a \times 2a \times 4c$. Despite this difference in out-of-plane modulation, the superstructure is described by the same space group, $Fddd$ \cite{PhysRevLett.129.137202}, and can be constructed using the layer stacking proposed in Ref.~\onlinecite{PhysRevB.84.104105}.

\begin{figure}
\centering
%\hspace{-5em}
\includegraphics[trim=10 100 550 135,clip,width=8.0cm]{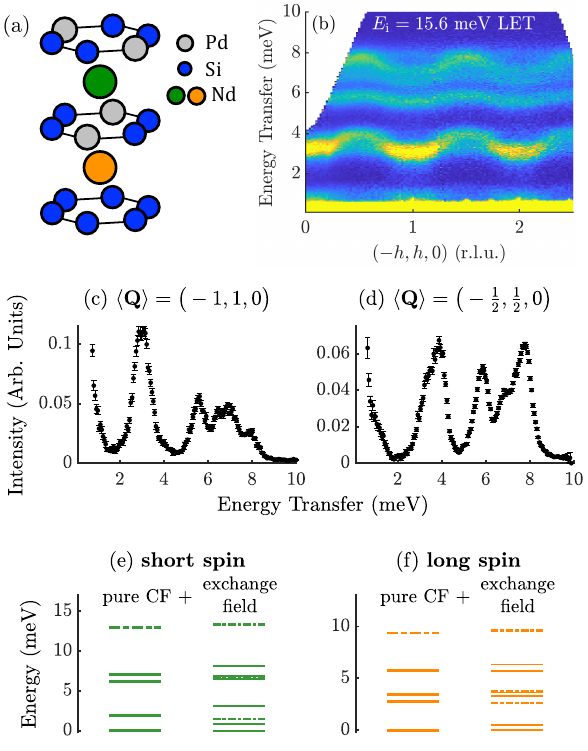}
\caption{(a) Part of the crystal structure of Nd$_{2}$PdSi$_{3}$, showing the local environments of the two Nd sites. Nd atoms (green/orange) may be 4-fold coordinated (top) or 2-fold coordinated (bottom) by Pd atoms. (b) Inelastic neutron scattering spectrum measured on LET along $(-h,h,0)$ at $1.7\,\mathrm{K}$ with neutrons of incident energy $E_\mathrm{i}=15.6\,\mathrm{meV}$. (c)--(d) Constant-$\mathbf{Q}$ cuts performed at $\langle\mathbf{Q}\rangle =(-1, 1, 0)$ and $(-\frac{1}{2}, \frac{1}{2}, 0)$~\textrm {r.l.u.}, respectively. Data were integrated over $\pm0.15~\textrm {r.l.u.}$ along $\langle\mathbf{Q}\rangle$ and two perpendicular directions. (e)--(f) Nd CF levels for (e) short-spin and (f) long-spin sites, according to our model. Continuous and dashed lines represent, respectively, levels with nonzero or zero INS cross section for excitation from the ground state. }\label{Fig1}
\end{figure}

The conventional unit cell of the Nd$_2$PdSi$_3$ superstructure contains a total of 32 Nd atoms divided equally among two inequivalent Nd sites. Both sites have point group symmetry $C_2$, but one is 2-fold and the other 4-fold coordinated by Pd [see Fig.~\hyperref[Fig1]{1(a)}]. Hence, the crystalline-electric field (CF) is expected to be different at these two sites. Assuming that the Nd atoms adopt the usual $3+$ valence, their $LS$-coupling $J=9/2$ ground-state multiplet is split by the orthorhombic CF into five Kramers doublets in the paramagnetic state. Below $T_\mathrm{C} = 14$--$17$\,K, when the compound develops a collinear ferromagnetic long-range order~\cite{KOTSANIDIS1990199,SZYTULA1999365,PhysRevB.68.012413,Xu2011,Mukherjee2011,PhysRevB.100.134423,suppl}, each doublet is further split into two singlets by the exchange field. For two sites, a maximum of $2\times (10-1)=18$ excited levels form the magnetic spectrum at sufficiently low temperatures. Details of the magnetic structure, determined here by neutron diffraction, are given in the Supplemental Material \cite{suppl}.

Figure~\hyperref[Fig1]{1(b)} shows the INS spectrum measured at a temperature of $1.7\,\mathrm{K}$ along the hexagonal reciprocal space direction $(-h,h,0)$ and Figs.~\hyperref[Fig1]{1(c)--(d)} are constant wavevector \textbf{Q} cuts performed for $h=1$ and $h=\frac{1}{2}$. Several dispersive modes, originating from the CF levels of the two inequivalent Nd sites, can be observed in the energy range from 2 to 8\,meV. Higher resolution data (see below) confirm that there are in fact three levels with measurable intensity at energies between 2.7 and 4\,meV, and another below 1\,meV. Consistent with Ref.~\onlinecite{PhysRevB.100.134423}, no modes were observed above 8\,meV, which implies that the spectrum manifest in Fig.~\hyperref[Fig1]{1(b)} represents the entire observable $J=9/2$ multiplet splitting of the Nd ions.

\begin{figure*}
\centering
\includegraphics[trim=0 240 0 40, clip,width=17.9cm]{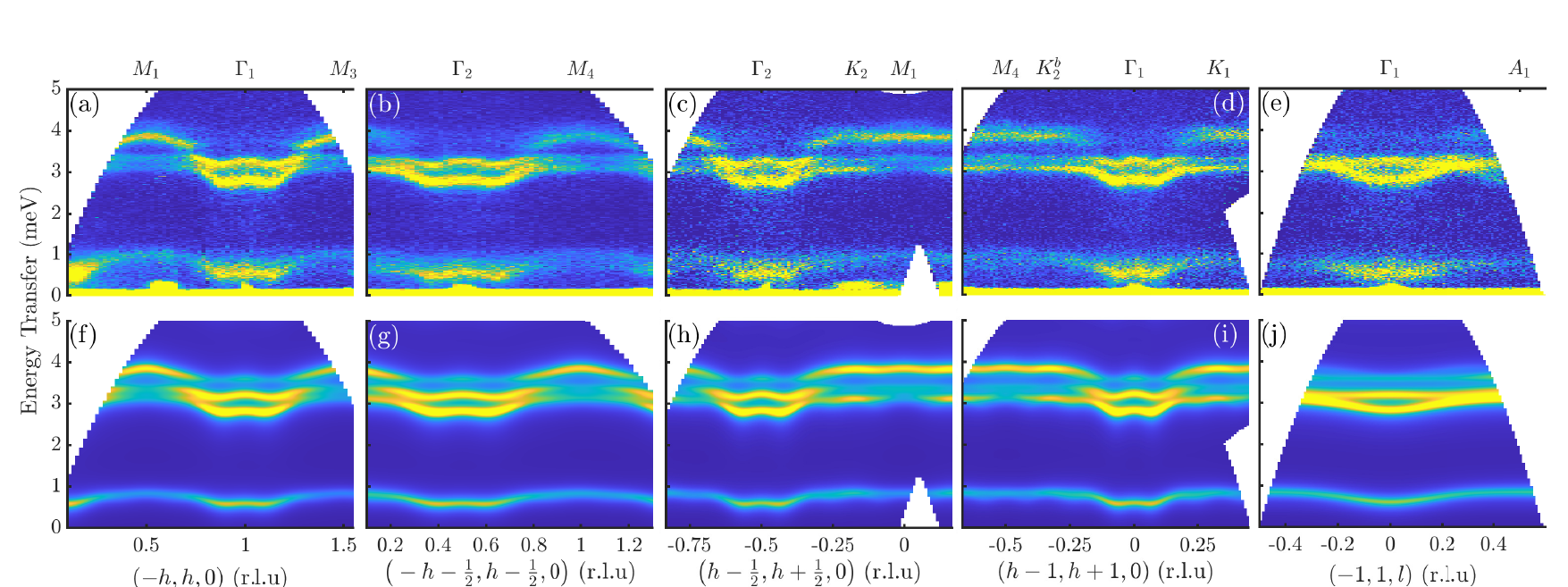}
\caption{(a)--(e) Spectra along several high-symmetry directions in reciprocal space measured on LET with neutrons of incident energy $E_\mathrm{i}=6\,\mathrm{meV}$, at a temperature of $1.7\,\mathrm{K}$. Brillouin zone high-symmetry labels are shown above the panels, as defined in Ref.~\onlinecite{suppl}. (f)--(j) Simulated spectra for the model which best describes the data displayed in (a)--(e). Positions in reciprocal space are indexed with respect to the parent $P6/mmm$ space group.}\label{fig2}
\end{figure*}

To model the data we used the mean-field random-phase approximation \cite{Jens_Mackintosh_book,Andrews_book}, which combines a description of the single-ion and two-ion interactions in a self-consistent way. The magnetic Hamiltonian is thus composed of two parts 
\begin{equation}
\mathcal{H} = \sum_i\mathcal{H}^\textup{CF}_{i} - \sum_{\langle ij \rangle} \mathbf{J}_i^{\mathrm{T}} \cdot \boldsymbol{\mathcal{J}}_{ij} 
%\textsf{J}_{ij}
\cdot \mathbf{J}_j,
\label{eq1}
\end{equation}
where $\mathcal{H}^\textup{CF}$ is the crystal field component and $\boldsymbol{\mathcal{J}}_{ij}$ are exchange matrices representing the (anisotropic) coupling between total angular momenta $\mathbf{J}_i$ and $\mathbf{J}_j$. We consider the approximate CF Hamiltonian 
\begin{equation}
\mathcal{H}^\textup{CF} = \sum_{k=2,4,6} B_0^k\hat{C}_0^{(k)}+B_6^6(\hat{C}_6^{(6)}+\hat{C}_{-6}^{(6)}),
\label{eq2}
\end{equation}
where $\hat{C}_{\pm q}^{(k)}$ are Wybourne tensor operators and $B_{q}^k$ the corresponding parameters. The site subscript is omitted here for simplicity. The Hamiltonian in Eq.~(\ref{eq2}) describes the leading-order $6/mmm$ point symmetry at the Nd site of the hexagonal parent structure. Although the true point symmetry requires more parameters, the truncated Hamiltonian (\ref{eq2}) already fully splits the $J = 9/2$ manifold. The higher order terms do not cause any further loss of degeneracy, so their effects can be approximately absorbed into the four parameters included in Eq.~(\ref{eq2}) and in the exchange Hamiltonian developed below. 

We find that a satisfactory fit can be achieved as long as a consistent assignment of levels is made to each one of the two Nd sites. Data collected in magnetic fields up to 11\,T applied along $\mathbf{c}$ were used to constrain the CF model \cite{suppl}. Figures~\hyperref[Fig1]{1(e)}--\hyperref[Fig1]{(f)} present the crystal-field scheme which best reproduces the observed excitations. Because $\mathcal{H}^\textup{CF}$ differs for both sites, the ground-state magnetic moments for the two Nd are distinct. The Nd with the lowest [highest] ground-state magnetic moment is referred to as \emph{short spin} [\emph{long spin}], see Fig.~\hyperref[Fig1]{1(e)} [\hyperref[Fig1]{1(f)}]. The $B_{q}^k$ parameters used to generate Figs.~\hyperref[Fig1]{1(e)}--\hyperref[Fig1]{(f)} are listed in Ref.~\onlinecite{suppl}. 

The determination of the exchange couplings was made using higher-resolution INS data shown in Figs.~\hyperref[fig2]{2(a)--(e)}, which provide a detailed view of the excitations up to 5~meV. Three levels can be resolved between 2 and 4~meV --- two of the modes soften at the $\Gamma$-point and, separated by a small gap, a third mode disperses to higher energies $\sim 4$~meV towards the M points. We identify these modes as three different CF levels. According to our model, the mode of lowest energy out of the trio belongs to the short-spin site, while the upper two belong to the long-spin site [see Figs.~\hyperref[Fig1]{1(e)--(f)}]. The level below 1\,meV originates from the splitting of the doublet ground-state on the short-spin sites. Although the ground state of the long-spin site also splits by $\sim$1\,meV, this transition has very small scattering intensity and cannot be observed in the data.

From the general softening of the modes near the $\Gamma$ point, see Figs.~\hyperref[fig2]{2(a)--(d)}, it can be inferred that the dominant nearest-neighbor in-plane couplings are ferromagnetic, which agrees with the magnetic structure reported in Ref.~\onlinecite{PhysRevB.100.134423} and also with our own neutron diffraction data \cite{suppl}. A striking additional feature of the in-plane dispersion is a high frequency modulation which has a local maximum at $\Gamma$. The periodicity of the oscillations, which is slightly different for the modes originating from each of the Nd sites, implies that higher-neighbor antiferromagnetic interactions are significant. We find that the closest in-plane antiferromagnetic coupling is between $5^\textrm{th}$-nearest neighbors, i.e.~$3^\textrm{rd}$-nearest neighbors within the $ab$ plane \cite{suppl}.

\begin{figure}
\centering
\subfigure{\includegraphics[trim=0 0 0 0, clip,width=7cm]{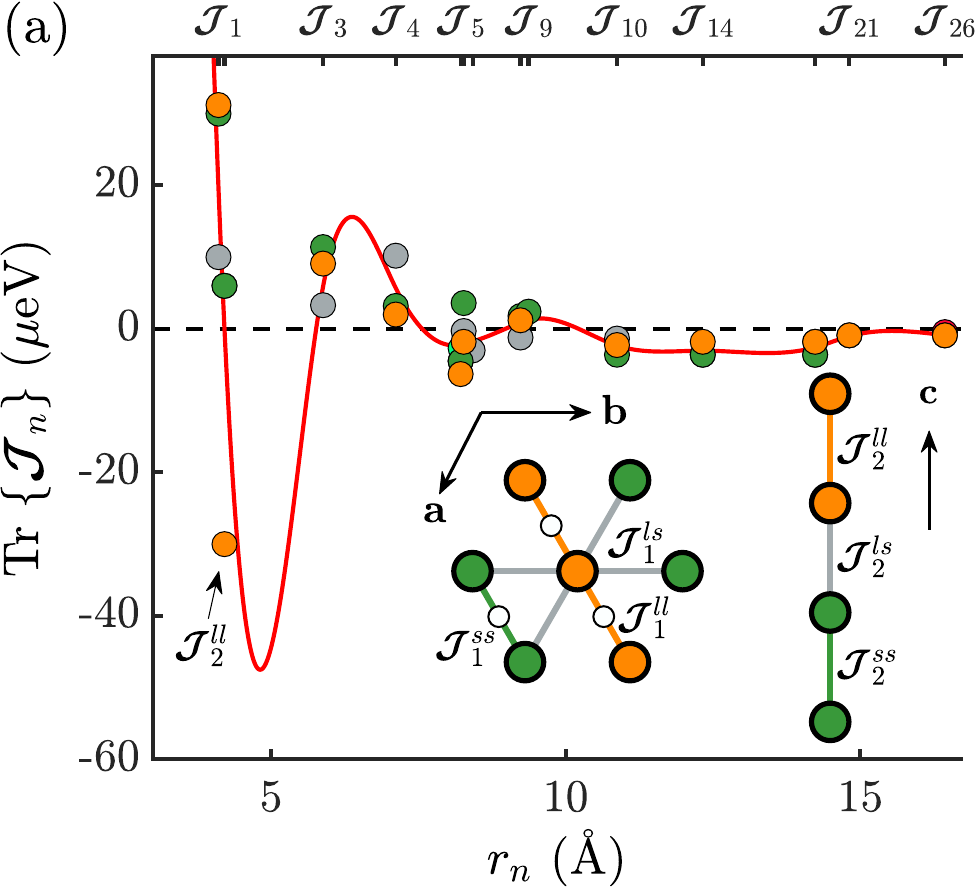}}
\subfigure{\includegraphics[trim=10 110 35 80, clip,width=8cm]{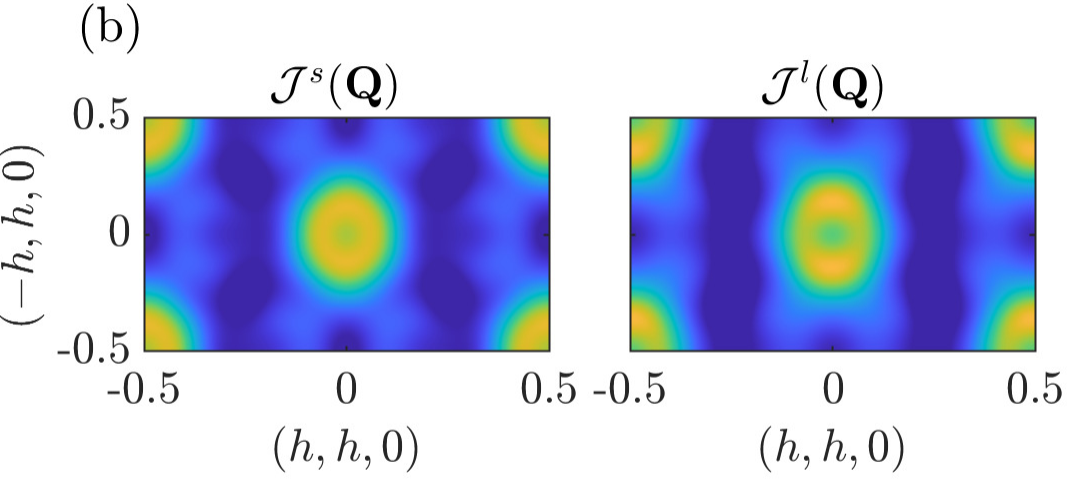}}
\caption{(a) Trace of the exchange coupling matrices used to obtain the spectra in Fig.~\ref{fig2}, as a function of $n^\textrm{th}$-neighbor distance $r_n$. Colors orange, gray and green correspond to $ll$, $ls$ and $ss$ couplings, respectively. The continuous line is a guide to the eye. Insets show first- and second-nearest neighbor exchange pathways, $\boldsymbol{\mathcal{J}}_{1}$ and $\boldsymbol{\mathcal{J}}_{2}$. Open circles indicate positions of inversion centers. (b) In-plane variation of the Fourier transform of $\textrm{Tr}\{\boldsymbol{\mathcal{J}}_{n}\}$ for each Nd site. Note: in hexagonal geometry the direction $(-h,h,0)$ is equivalent to $(h,0,0)$.}\label{fig3}
\end{figure}

The simulated spectra shown in Figs.~\hyperref[fig2]{2(f)--(j)} are from the best model we could find to describe the data, taking into account the six possible superstructure domains \cite{suppl}.
The simulations reproduce all the main observed features. Figure~\hyperref[fig3]{3(a)} plots the model exchange parameters as represented by the trace of the exchange matrices as a function of the $n^\textrm{th}$-neighbor distance $r_n$. The exchange interactions are seen to extend to large distances and tend to oscillate with $r_n$. With two distinct Nd sites in the superstructure, symmetry-inequivalent exchange couplings cannot be discriminated simply by bond distance. Because of that, the $\boldsymbol{\mathcal{J}}_{ij}$ in Fig.~\hyperref[fig3]{3(a)} are labeled with a superscript --- long-long ($ll$), short-short ($ss$) or long-short ($ls$) --- indicating which sites are coupling, and a subscript $n$ for bonds at a distance $r_n$.   These are illustrated in the inset of Fig.~\hyperref[fig3]{3(a)} for first and second neighbors. A complete symmetry analysis of the couplings and the assumptions made in this work can be found in the Supplemental Material~\cite{suppl}. Figure~\hyperref[fig3]{3(b)} shows the Fourier transform $\mathcal{J}(\mathbf{Q})$ of the in-plane exchange interactions calculated for each Nd site from the values of $\textrm{Tr}\{\boldsymbol{\mathcal{J}}_{n}\}$. The maximum in $\mathcal{J}(\mathbf{Q})$ is found at the hexagonal reciprocal lattice vector $\mathbf{Q}_\textrm{m} = (q,0,0)$, where $q \simeq 0.12$/0.14 for the \textit{s}/\textit{l} sites, respectively.

Figure~\hyperref[fig2]{2(e)}, which shows the out-of-plane dispersion measured along $(-1,1,l)$, reveals something else about the exchange interactions.  The modes originating from the short-spin sites (centered on $\sim$ 0.8 and 3\,meV) are seen to soften by about 0.4\,meV at the $\Gamma_1$ point, whereas the modes associated with the long-spin sites (3.3 and 3.6\,meV) are essentially dispersionless. The observation of flat modes suggests that the magnetic coupling between the long-spin sites is highly frustrated. Consistent with this, our model finds the nearest-neighbor in-plane and out-of-plane couplings between long spins to be similar in magnitude but opposite in sign, as can be seen from the points (orange symbols) in Fig.~\hyperref[fig3]{3(a)} representing $\boldsymbol{\mathcal{J}}^{ll}_1$ and $\boldsymbol{\mathcal{J}}^{ll}_2$. By contrast, we find no significant short-range frustration within the Nd layers, e.g.~the $\boldsymbol{\mathcal{J}}_5$ and $\boldsymbol{\mathcal{J}}_1$ couplings also have opposite sign but differ by an order of magnitude.

\begin{figure}
\centering
\includegraphics[trim=5 10 28 10, clip,width=8.4cm]{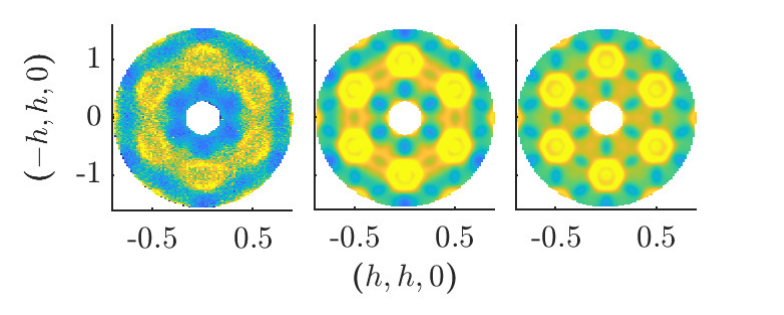}
\caption{Constant-energy slices through data (left) and models (middle and right) integrated over the energy interval 3.0 to $3.3\,\mathrm{meV}$. The $360^\circ$ data map was constructed by rotation of a measured $120^\circ$ sector. Middle panel shows simulations using anisotropic exchange coupling, as in Figs.~\hyperref[fig2]{2(f)--(j)}. Right panel is for an isotropic exchange model.}\label{fig4}
\end{figure}

Our simplifying assumption of 6-fold rotational symmetry for $\mathcal{H}^\mathrm{CF}$, Eq.~(\ref{eq2}), constrains the single-ion anisotropy to be isotropic in the $ab$ plane. The true $Fddd$ symmetry, however, allows in-plane single-ion anisotropy as well as anisotropic $\boldsymbol{\mathcal{J}}_{ij}$ couplings along most paths \cite{suppl}. Although not apparent in Fig.~\ref{fig2}, the effects of magnetic anisotropy can be seen in the constant-energy maps of Fig.~\ref{fig4}, which integrate over the middle of the three bands located between 3 and 4\,meV. The data (left panel) display two-fold symmetry around the reciprocal lattice points, whereas the intensity calculated from a model with isotropic exchange interactions (right panel) has six-fold symmetry. The main features of the data are reproduced reasonably well by our model with anisotropic exchange interactions (middle panel). The intensity map has six-fold symmetry around the origin because of the averaging of the six two-fold symmetric domains. We stress that anisotropy in the spectrum could also arise from higher-order terms in $\mathcal{H}^\mathrm{CF}$, but the effects of single- and two-ion anisotropy are very difficult to separate.

\begin{figure}
\centering
\includegraphics[trim=0 0 0 0, clip,width=8.2cm]{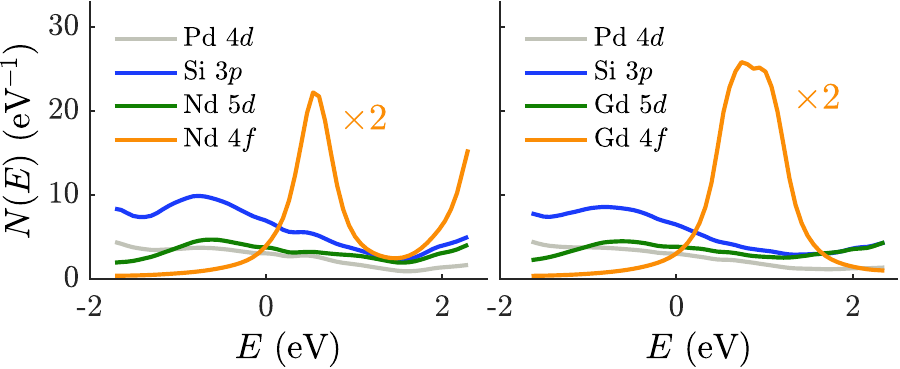}
\caption{Calculated density of states for selected bands close to the Fermi energy ($E=0$\,eV) for Nd$_2$PdSi$_3$ (left) and Gd$_2$PdSi$_3$ (right). The Nd$_2$PdSi$_3$ superstructure is assumed in both cases. The $4f$ bands have been scaled down by a factor of two. }\label{fig5}
\end{figure}

Our model has a large parameter set, and there may exist other sets which give similar agreement. However, by exploring the effects of varying the parameters we have found that the general features summarised above are robust. Moreover, there are good reasons to expect that our overall findings will extend to the SkX compound Gd$_2$PdSi$_3$. First, the superstructures of the Nd and Gd compounds are essentially the same (the only difference being a doubling along the $\mathrm{c}$ axis in the latter). Second, our first-principles calculations (see Fig.~\ref{fig5} and Supplemental Material \cite{suppl}), as well as experimental studies \cite{Inosov2009}, indicate that the electronic structure near the Fermi energy in $R_2$PdSi$_3$ does not vary significantly with $R$. In particular, the $5d$ bands, which are believed to be important for skyrmion formation \cite{PhysRevLett.125.117204}, are very similar in the two compounds (Fig.~\ref{fig5}). Third, the $\mathcal{J}(\mathbf{Q})$ for Nd$_2$PdSi$_3$, Fig.~\hyperref[fig3]{3(b)}, is maximal at a wave vector  with the same direction and similar magnitude $(q = 0.12$ to $0.14)$ as the observed zero-field magnetic propagation vector of Gd$_2$PdSi$_3$ $(q \simeq 0.14)$~\cite{Kurumaji_science}. This implies that the magnetic interactions in the Nd and Gd compounds are similar. Incommensurate in-plane magnetic order does not occur in the Nd case because the crystal-field  anisotropy locks the Nd moments along the $c$ axis and drives ferromagnetic order.

Consequently, our comprehensive study of the magnetic spectrum of single-crystal Nd$_2$PdSi$_3$ has important implications for the formation of a SkX phase in Gd$_2$PdSi$_3$. Our data reveal a critical dependence on the Pd/Si superstructure, such that the pattern of magnetic interactions has orthorhombic, not hexagonal symmetry. Our results also raise questions about the role played by geometric frustration in the magnetism displayed by this family of compounds. Although there are some frustrated magnetic interactions, especially associated with the antiferromagnetic $\boldsymbol{\mathcal{J}}^{ll}_2$ nearest-neighbor coupling along the $c$ axis, we do not find evidence for strong short-range frustration within the Nd layers. Our analysis, moreover, has revealed the long-range nature of the magnetic interactions, with non-negligible couplings extending up to at least the $26^\textrm{th}$ nearest neighbor and a tendency to oscillate with distance. The evidence presented here suggests that theories for the SkX phase in Gd$_2$PdSi$_3$ must go beyond simple models of short-range frustrated interactions on a triangular lattice, and point instead towards approaches based on longer-range RKKY exchange. Such considerations may also apply to other metallic centrosymmetric SkX materials.

\section*{Acknowledgments}
The research at Oxford was supported by UK Research and Innovation, grant No.~EP/T027991/1, and by the Oxford--ShanghaiTech collaboration project. The work at Warwick was supported by UK Research and Innovation, grant Nos.~EP/T005963/1 and EP/N032128/1. J.B.~was supported by the Alexander von Humboldt Foundation through the Feodor Lynen Research Fellowship for Postdocs. Data from the neutron experiments are available from the STFC ISIS Neutron and Muon Source, proposal Nos.~RB2010616 (Merlin \cite{MS_Merlin_2020}), RB2000180 (Merlin \cite{MS_Merlin_2020_2}), and RB2310234 (LET \cite{VPA_LET_2023}), and from the Institut Laue-Langevin, proposal No.~5-41-1185 (D3 \cite{VPA_D3_2023}). Other data are available from the authors upon reasonable request. 

\bibliography{./bibs/PhysRevB.84.104105, 
./bibs/PhysRevB.100.134423,
./bibs/Kurumaji_science,
./bibs/Jens_Mackintosh_book,
./bibs/PhysRevLett.108.017206,
./bibs/PhysRevLett.124.207201,
./bibs/Leonov2015,
./bibs/Muhlbauer,
./bibs/Yu,
./bibs/Fert2013,
./bibs/Lancaster2019,
./bibs/Bogdanov2020,
./bibs/Takagi2022,
./bibs/PhysRevLett.102.197202,
./bibs/Wiesendanger,
./bibs/PhysRevLett.127.067201,
./bibs/Moreau-Luchaire,
./bibs/Thomas,
./bibs/PhysRevX.9.041063,
./bibs/PhysRevLett.130.106701,
./bibs/Mallik_1998,
./bibs/KOTSANIDIS1990199,
./bibs/MALLIK1998169,
./bibs/PhysRevB.62.425,
./bibs/Frontzek_2007,
./bibs/FRONTZEK2006398,
./bibs/PhysRevB.68.012413,
./bibs/PhysRevB.62.14207,
./bibs/Nentwich_2016,
./bibs/SZYTULA1999365,
./bibs/PhysRevB.64.012418,
./bibs/PhysRevB.60.12162,
./bibs/PhysRevB.67.212401,
./bibs/Hirschberger_2019,
./bibs/Khanh_2020,
./bibs/Gao_2020,
./bibs/GORDON199724,
./bibs/CHEVALIER1984753,
./bibs/GLADYSHEVSKII1992221,
./bibs/PhysRevB.103.104408,
./bibs/PhysRevLett.125.117204,
./bibs/PhysRevB.95.224424,
./bibs/PhysRevB.93.064430,
./bibs/PhysRevLett.129.137202,
./bibs/Andrews_book,
./bibs/PhysRevB.104.184432,
./bibs/PhysRevLett.128.157206,
./bibs/PhysRevLett.133.016401,
./bibs/Mukherjee2011,
./bibs/Xu2011,
./bibs/suppl,
./bibs/PhysRevB.103.024439,
./bibs/MS_Merlin_2020,
./bibs/MS_Merlin_2020_2,
./bibs/VPA_LET_2023,
./bibs/VPA_D3_2023,
./bibs/Ju2023,
./bibs/Inosov2009,
./bibs/PhysRevLett.134.046702,
./bibs/PhysRevB.105.054435}
\end{document}

\title{Supplemental Material for ``Long-range magnetic interactions in Nd$_2$PdSi$_3$ and the formation of skyrmion phases in centrosymmetric metals''}

\author{Viviane~Pe\c{c}anha-Antonio}
\email{viviane.antonio@stfc.ac.uk}
\affiliation{Department of Physics, University of Oxford, Clarendon Laboratory, Oxford OX1 3PU, United Kingdom}
\affiliation{ISIS Neutron and Muon Source, STFC Rutherford Appleton Laboratory, Didcot OX11 0QX, United Kingdom}
\author{Zhaoyang Shan}
\author{Michael~Smidman}
\affiliation{Center for Correlated Matter and School of Physics, Zhejiang University, Hangzhou 310058, China}
\author{Juba Bouaziz}
\affiliation{Department of Physics, The University of Tokyo, Bunkyo-ku, Tokyo, 113-0033, Japan}
\author{Bachir~Ouladdiaf}
\author{Iurii~Kibalin}
\author{Marie-H\'{e}l\`{e}ne~Lem\'{e}e-Cailleau}
\affiliation{Institut Laue-Langevin, 6 Rue Jules Horowitz, BP 156, 38042 Grenoble Cedx 9, France}
\author{Christian~Balz}
\affiliation{ISIS Neutron and Muon Source, STFC Rutherford Appleton Laboratory, Didcot OX11 0QX, United Kingdom}
\author{Jakob~Lass}
\affiliation{PSI Center for Neutron and Muon Sciences, 5232 Villigen PSI, Switzerland}
\author{Daniel~A.~Mayoh}
\author{Geetha~Balakrishnan}
\author{Julie B. Staunton}
\affiliation{Department of Physics, University of Warwick, Coventry CV4 7AL, United Kingdom}
\author{Devashibhai~Adroja}
\affiliation{ISIS Neutron and Muon Source, STFC Rutherford Appleton Laboratory, Didcot OX11 0QX, United Kingdom}
\affiliation{Highly Correlated Matter Research Group, Physics Department, University of Johannesburg, Auckland Park 2006, South Africa}
\author{Andrew~T.~Boothroyd}
\email{andrew.boothroyd@physics.ox.ac.uk}
\affiliation{Department of Physics, University of Oxford, Clarendon Laboratory, Oxford OX1 3PU, United Kingdom}

\date{\today}

\maketitle

\section{Experimental methods}

A single crystal of Nd$_2$PdSi$_3$ was grown by the optical floating-zone method, see Ref.~\onlinecite{MAYOH2024127774} for details. The grown crystal was in the form of a rod of approximate dimensions 5\,mm (diameter) $\times$ 50\,mm (length). Parts of the same rod were used for all the measurements in this work. X-ray single-crystal diffraction was carried out at room temperature on a SuperNova Agilent diffractometer equipped with a Mo source.

\begin{figure}
\centering
\includegraphics[trim=0 0 0 0, clip,width=7cm]{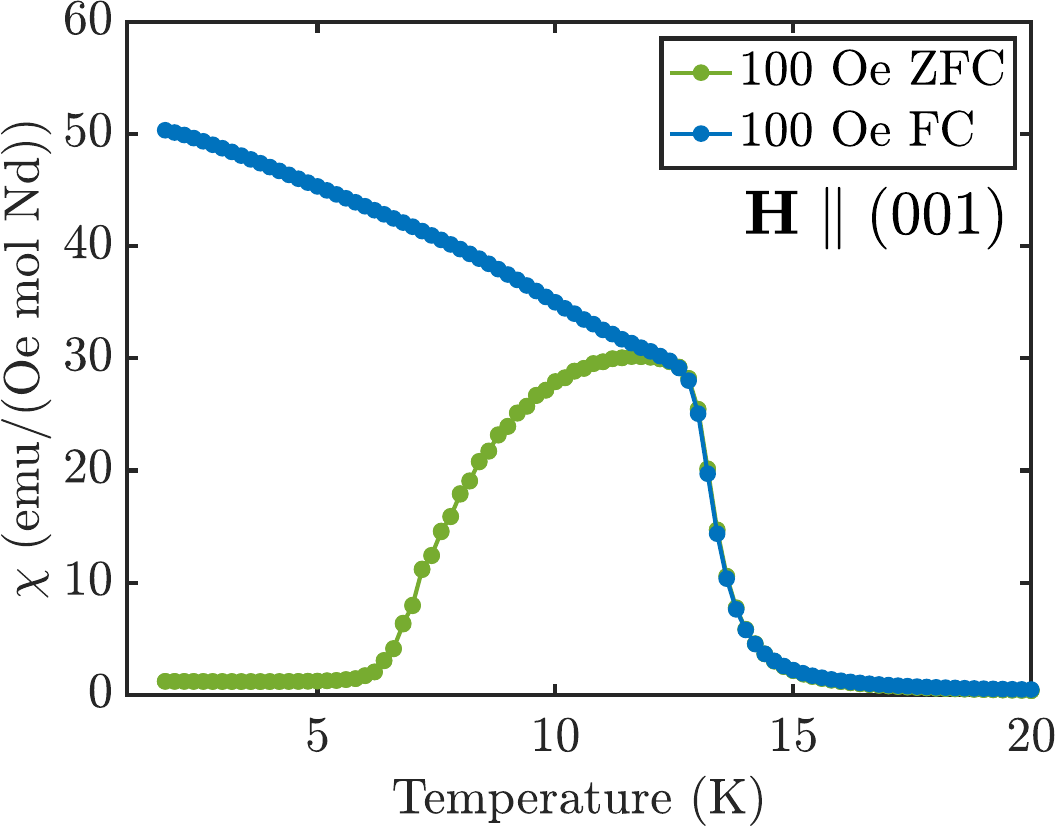}
\caption{Magnetic susceptibility measured on a single crystal of Nd$_2$PdSi$_3$ with $\mathbf{H} \parallel \mathbf{c}$. The sharp increase in susceptibility around 14\,K marks the onset of ferromagnetism.}\label{Susceptibility}
\end{figure}

Magnetic susceptibility measurements were performed with a Quantum Design MPMS3 SQUID-VSM magnetometer. The same piece of crystal was used for the neutron diffraction and magnetometry measurements. Fig.~\ref{Susceptibility} shows susceptibility data for a magnetic field of 100\,Oe applied along the crystallographic $\mathbf{c}$ axis. The onset of spontaneous magnetisation in the sample takes place at $T_\mathrm{C}\simeq 14$\,K. This temperature is about 1.7\,K lower than that reported for powder samples in Refs.~\onlinecite{SZYTULA1999365,PhysRevB.100.134423}.   

Neutron diffraction measurements were performed on the four-circle diffractometer D10+ at the Institut Laue--Langevin (ILL). A single crystal of dimensions $2\,\mathrm{mm}\times1\,\mathrm{mm}\times2\,\mathrm{mm}$ was cut from the main rod and fixed on a standard aluminium mount which was installed in a four-circle cryostat. Neutrons of wavelengths $\lambda = 1.26$\,\AA~and $2.36$\,\AA~were selected by Bragg diffraction from a Cu monochromator and a pyrolytic graphite monochromator, respectively. A graphite filter was used to suppress higher order harmonic wavelengths in the incident beam. The scattered neutrons were recorded on a $94 \times
94$\,mm$^2$ area detector. In the paramagnetic phase, a total of 740 inequivalent reflections were recorded using $\lambda = 1.26$\,\AA~at a temperature of 150\,K. Using $\lambda = 2.36$\,\AA, 111 inequivalent reflections were measured at 20\,K, just above the phase transition temperature, and 94 at 2\,K.

For the inelastic neutron scattering (INS) experiments, a sample of total mass around $5$\,g was fixed in an aluminium plate with the hexagonal $[0 0 1]$ axis vertical. Time-of-flight INS was performed on the MERLIN and LET chopper spectrometers \cite{MERLIN,LET} at the ISIS Facility. On MERLIN, the experiment was divided in two parts making use of two different sample environments. In the first part, the sample was loaded in a closed-cycle refrigerator (CCR) and data were recorded at 7.7\,K and 30\,K. In the second, the sample was loaded in a helium cryostat and data were collected at 1.8\,K. For both parts, the crystal was rotated through an angle of $125^\circ$ in $1^\circ$ steps around the vertical axis. The instrument chopper frequency was set at 250\,Hz and operated in repetition-rate multiplication (RRM) mode, so that angular scans with incident energies $E_\textrm{i}$ of 10, 20 and 53\,meV were performed simultaneously. 

On LET \cite{LET}, the single crystal was fixed in an aluminium mount and loaded in a helium cryostat. During data collection, at a temperature of 1.8\,K, the sample was rotated through an angle of $120^\circ$ in $1^\circ$ steps around the vertical axis. RRM enabled the simultaneous measurement of $E_\textrm{i}=6$ and 15.6\,meV, among other incident energies with less flux. For the former energy transfer, the FWHM of the energy resolution at the elastic line is 0.16\,meV, down to about 0.06\,meV at 5\,meV. 

Inelastic neutron scattering data in magnetic field were collected on the cold neutron multiplexing spectrometer CAMEA \cite{CAMEA} at the Swiss Spallation Neutron Source at the Paul Scherrer Institut (PSI). For all magnetic fields, measurements were performed with $E_\mathrm{i}=7$ and $8$\,meV ($+0.13$ meV for interlacing) and with the centre of the detector tank at angles of $2\theta = -44^\circ$ and $-48^\circ$. In each case, the sample was rotated through an angle of 141$^\circ$ in $1^\circ$ steps. For 0, 4, and 7\,T also the elastic line was measured, i.e. $E_\mathrm{i}=5$ and $5.13\,\mathrm{meV}$ at $2\theta = -51^\circ$ and $-56^\circ$. In addition, at 0 and 7\,T the spectrum was measured up to 11\,meV with $E_\mathrm{i} = 10,\,10.13,\,12,\,12.13,\,14,\,14.13$ and $2\theta = -38^\circ$ and $-42^\circ$, covering $40^\circ$ of sample rotation in steps of $1^\circ$. All points were measured with a monitor of 250 000 corresponding to $\sim 1\,\mathrm{minute}$ at $E_\mathrm{i}=5\,\mathrm{meV}$ and 2 minutes at $E_\mathrm{i}=14\,\mathrm{meV}$, except for the 0\,T dataset, where a 190 000 monitor was used. Data were analysed using the MJOLNIR software package \cite{MJOLNIR}.

\section{Crystallographic superstructure and static magnetism}

\begin{figure}
\centering
\includegraphics[trim=30 230 60 250, clip,width=8.5cm]{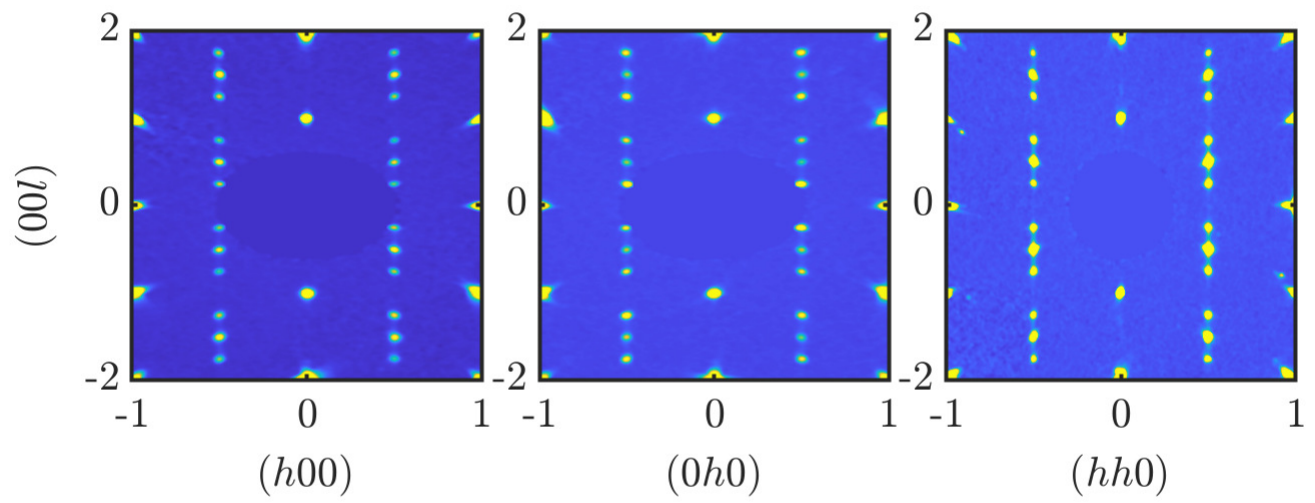}
\caption{X-ray diffraction in the $(h0l)$, $(0hl)$ and $(hhl)$ reciprocal lattice planes, measured at room temperature.}\label{HHL}
\end{figure}

The superstructure of the heavy rare-earth $R_2$PdSi$_3$ series ($R=$ Gd, Tb, Dy, Ho, Er, Tm) was thoroughly studied in the work of Tang \emph{et al.} \cite{PhysRevB.84.104105}. In those systems, the unit cell is doubled along the hexagonal \textbf{a} and \textbf{b} directions and octupled along \textbf{c}. Our single-crystal x-ray diffraction measurements, represented in Fig.~\ref{HHL}, demonstrate that in Nd$_2$PdSi$_3$ the superstructure unit cell is enlarged by a factor $2a\times2a\times4c$ instead. Peaks are observed in the $(hhl)$ reciprocal lattice plane at positions with $h = (2n+1)/2$ and $l = m/4$, $l\ne$ integer, where the indices $h$, $k$ and $l$ refer to the parent hexagonal lattice and $n,m$ are integers. The same conditions on $h$ and $l$ apply for the $(h0l)$ and $(0hl)$ reciprocal lattice planes. 

\begin{figure}
\centering
\includegraphics[trim=0 0 0 0, clip,width=8.5cm]{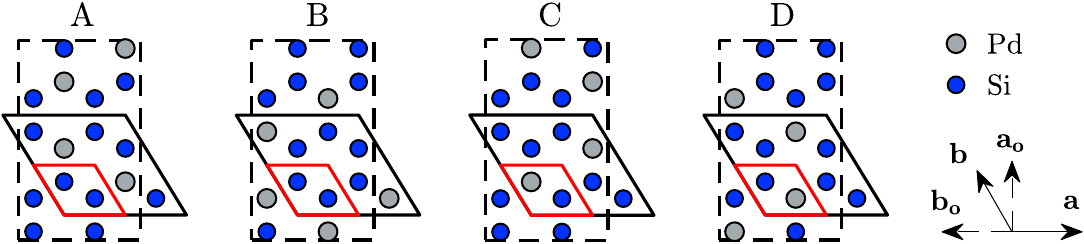}
\caption{Layers of ordered Pd and Si atoms in the superlattice $ab$ plane of Nd$_{2}$PdSi$_{3}$. The four layers, labeled A, B, C and D, are stacked along the $\mathbf{c}$ axis according to the sequences listed in Table~\ref{tab1} to form the different superstructure domains. Dashed and continuous black lines indicate the orthorhombic and hexagonal superstructure unit cells, respectively. The corresponding cell vectors are $\textbf{a}_\textrm{o}$, $\textbf{b}_\textrm{o}$ and $\textbf{a}$, $\textbf{b}$. The unit cell of the parent \emph{P6/mmm} space group is shown in red.}\label{Layers}
\end{figure}

The superstructure proposed in Ref.~\onlinecite{PhysRevB.84.104105} is formed by the stacking of four layers in which the Pd atoms have distinct positions. These layers are labeled A, B, C and D, and are shown in Fig.~\ref{Layers}. In order to explain the extinction rules observed in their diffraction data, one of the assumptions made by the authors of Ref.~\onlinecite{PhysRevB.84.104105} is that no two adjacent layers along the $\mathbf{c}$ axis are of the same type. Given that the periodicity of the unit cell in Nd$_2$PdSi$_3$ is vertically increased by a factor of 4, the minimal unit cell which can be hypothesised in our case contains a vertical stacking of all the four layers shown in Fig.~\ref{Layers}. 

Following symmetry considerations, it can be shown that six different twin domains contribute to the measured structure factors. These domains can be found by applying the symmetry operations $\{ 1 , 3^{-}_{001}, 3^{+}_{001}, \bar{1}, \bar{3}^{-}_{001}, \bar{3}^{+}_{001}$ of the \emph{P6/mmm} space group to the stacking ABCD, which we refer to as \emph{Domain 1}. The particular layer stacking of each domain is given in Table~\ref{tab1}. In Fig.~\ref{Layers}, we reproduce the layers A, B, C and D in the orthorhombic supercell, and show the relation between these lattice basis vectors with the hexagonal cell. Transformation matrices ${\mathbf{T}}_m$ relating the coordinates of domains $m=2,3,...,6$ with domain 1 in the $Fddd$ space group are given below. The notation used combines a $ 3 \times 3 $ rotation matrix with a translation vector in the last column.

\begin{align}
{\mathbf{T}}_{2}=
\begin{pmatrix}
-1/2 \ & \ -1/2 \ & \ 0 \ & \ 1/2\\
3/2 \ & \ -1/2 \ & \ 0 \ & \ 3/4 \\
0 \ & \ 0 \ & \ 1 \ & \ 0 \\
\end{pmatrix}.\label{eqT2}
\end{align}

\begin{align}
{\mathbf{T}}_{3}=
\begin{pmatrix}
-1/2 \ & \ 1/2 \ & \ 0 \ & \ -1/8\\
-3/2 \ & \ -1/2 \ & \ 0 \ & \ 9/8 \\
0 \ & \ 0 \ & \ 1 \ & \ 0 \\
\end{pmatrix}.
\end{align}

\begin{align}
{\mathbf{T}}_{4}=
\begin{pmatrix}
-1 \ & \ 0 \ & \ 0 \ & \ 1/4\\
0 \ & \ -1 \ & \ 0 \ & \ 5/4 \\
0 \ & \ 0 \ & \ -1 \ & \ 0 \\
\end{pmatrix}.
\end{align}

\begin{align}
{\mathbf{T}}_{5}=
\begin{pmatrix}
1/2 \ & \ 1/2 \ & \ 0 \ & \ -1/4\\
-3/2 \ & \ 1/2 \ & \ 0 \ & \ 1/2 \\
0 \ & \ 0 \ & \ -1 \ & \ 0 \\
\end{pmatrix}.
\end{align}

\begin{align}
{\mathbf{T}}_{6}=
\begin{pmatrix}
1/2 \ & \ -1/2 \ & \ 0 \ & \ 3/8\\
 3/2 \ & \ 1/2 \ & \ 0 \ & \ 1/8 \\
 0 \ & \ 0 \ & \ -1 \ & \ 0 \\
\end{pmatrix}.\label{eqT6}
\end{align}

\begin{table}
\begin{ruledtabular}
\begin{tabular}{l r}
domain & Stacking \\
\colrule
1& ABCD\\
2& ADBC\\
3& ACDB \\
4& DCBA\\
5& CBDA\\
6& BDCA\\
\end{tabular}
\end{ruledtabular}
\caption{The layer stacking of the six domains obtained by applying the symmetry operations $\{ 1 , 3^{-}_{001}, 3^{+}_{001}, \bar{1}, \bar{3}^{-}_{001}, \bar{3}^{+}_{001}$ of the \emph{P6/mmm} space group to the ABCD sequence of planes. An alternative way of generating the six domains is considering all the possible ways of stacking vertically A, B, C and D, requiring that no adjacent layer is the same and neglecting cyclic plane permutations, which are necessarily equivalent.}\label{tab1}
\end{table}

\subsection{Structure refinement}

The x-ray data measured at room temperature, and the neutron diffraction data measured in the paramagnetic phase at 150\,K, were refined in the \emph{Fddd} group description using FullProf \cite{CARVAJAL1}. The model assumes an equal population of the six domains. Atomic positions of the general site ($32h$) were not refined. In the reciprocal space maps shown in Fig.~\ref{HHL}, a weak diffuse scattering can be observed along the $(00l)$ direction, evidencing that stacking faults occur. In order to (partially) account for that in the refinement of the peak intensities, the average occupation of Pd and Si sites was initially allowed to vary equally for all the domains, while the overall stoichiometry, i.e.~three atoms of Si for each Pd, was kept constrained. Isotropic displacement parameters were refined at a later stage, during which the occupancies were kept constant at a value which gave the best agreement with the measured patterns.

\begin{figure}
\centering
\includegraphics[trim=0 0 0 0, clip,width=8.5cm]{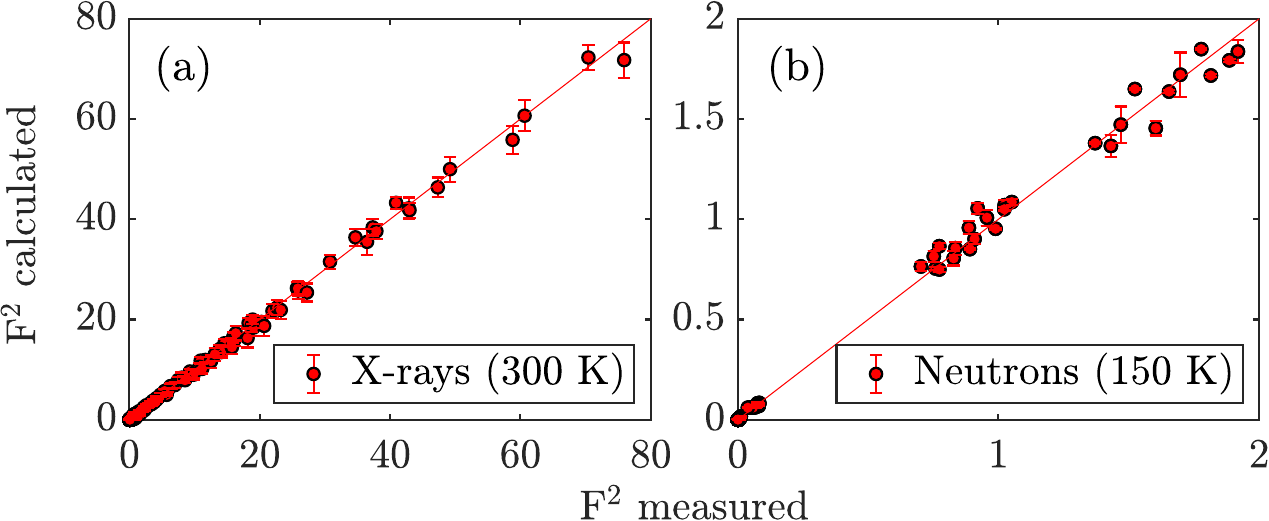}
\caption{Refinement results of the single crystal diffraction measured with (a) x-rays at room temperature, and (b) neutrons on D10+ at $150\,\mathrm{K}$.}\label{refinement_structure}
\end{figure}

A summary of the x-ray diffraction refinement results for domain 1 is presented in Table~\ref{tab2}. A quantitative comparison between calculated and measured structure factors is shown in Figs.~\hyperref[refinement_structure]{4(a)} and \hyperref[refinement_structure]{4(b)} for x-rays and neutrons, respectively. 

\begin{table*}
\begin{ruledtabular}
\begin{tabular}{l c c c}
Lattice parameters & $\mathbf{a}_\textrm{o}=14.2402$\,\AA & $\mathbf{b}_\textrm{o}=8.2216$ \,\AA & $\mathbf{c}_\textrm{o}=16.8595$\,\AA \\
Lattice angles & $\alpha=90^\circ$ & $\beta=90^\circ$ & $\gamma=90^\circ$ \\
\colrule
Atom & Atomic position & Occupancy & Isotropic displacement \\
& (Wyckoff site)& & parameters (\AA$^2$)\\
\colrule
Nd& 1/8,5/8,0 (16g) & $0.5$ & 0.52(3) \\
Nd& 1/8,5/8,1/2 (16g) & $0.5$ & 0.52(3) \\
Pd/Si&  7/24,1/8,1/8 (16e) & $0.60(1)/0.4(1)$ & 0.64(4)\\
Si/Pd& 11/24,1/8,1/8 (16e) & $0.866(1)/0.134(1)$ & 0.64(4)\\
Si/Pd& 13/24,7/8,1/8 (32h) & $0.867(1)/0.133(1)$ & 0.64(4)\\
\end{tabular}
\end{ruledtabular}
\caption{Summary of the structural model assumed for the superstructure of Nd$_2$PdSi$_3$ in the space group $Fddd$. The atomic positions in the other five domains can be obtained with the transformation matrices ${\mathbf{T}}_m$ defined in Eqs.~(\ref{eqT2})-(\ref{eqT6}) above.}\label{tab2}
\end{table*}

\subsection{Magnetic structure refinement}

\begin{figure}
\centering
\includegraphics[trim=0 0 0 0, clip,width=8.5cm]{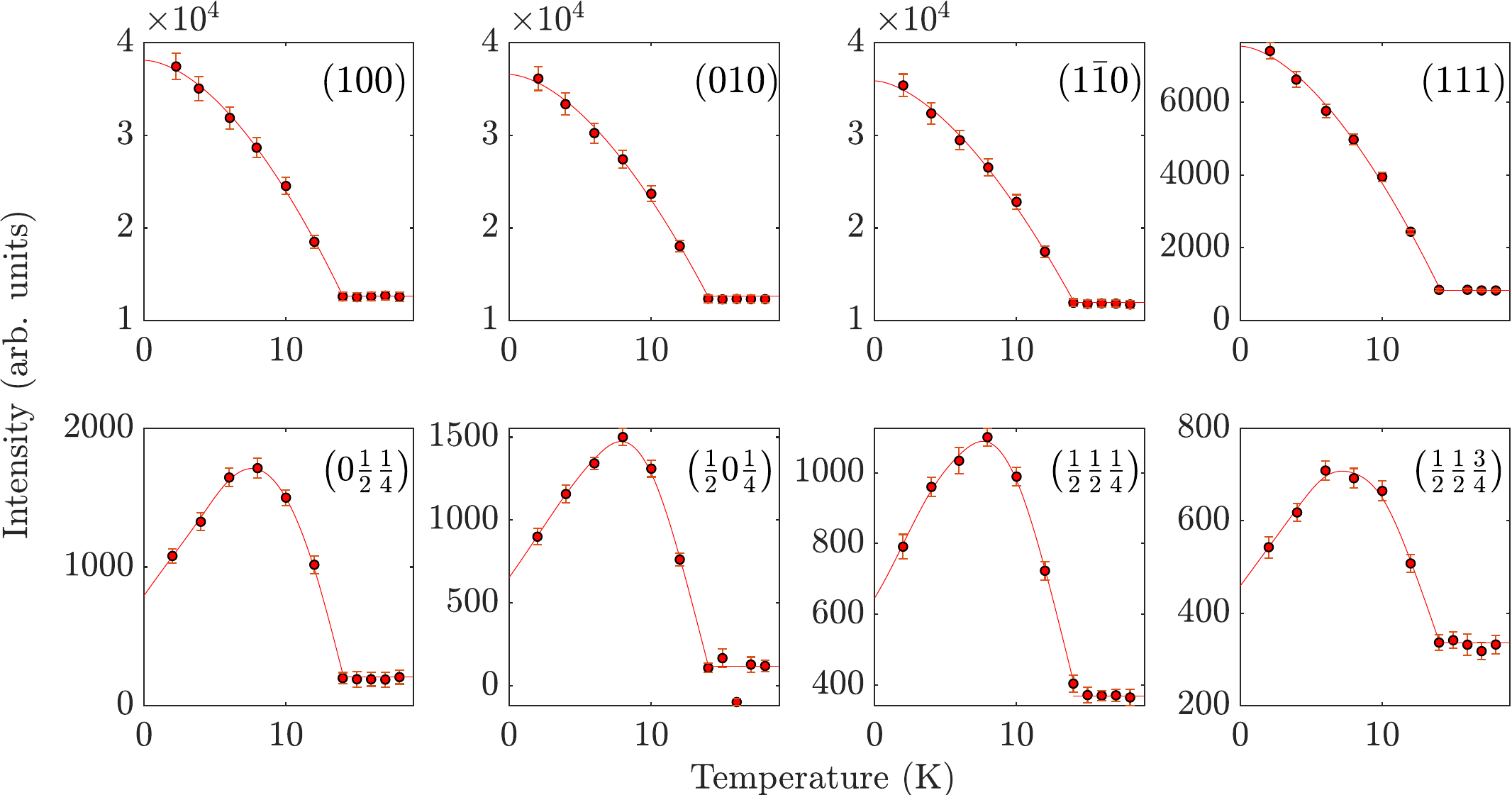}
\caption{Temperature dependence of Bragg-peak intensities for parent structure and superstructure reflections. For clarity, the indexing follows the hexagonal $\emph{P6/mmm}$ space group description. Red lines are a guide to the eye.}\label{T_dependence}
\end{figure}

It was found in Ref.~\onlinecite{PhysRevB.100.134423} that, in the magnetic long-range ordered phase, magnetic scattering is detected in integer- and non-integer-indexed Bragg peaks, and the temperature dependence of these two sets is different. As expected, a similar behaviour is observed in our single-crystal neutron diffraction results, as may be seen in Fig.~\ref{T_dependence}. 

\begin{figure}
\centering
\includegraphics[trim=0 0 0 0, clip,width=7cm]{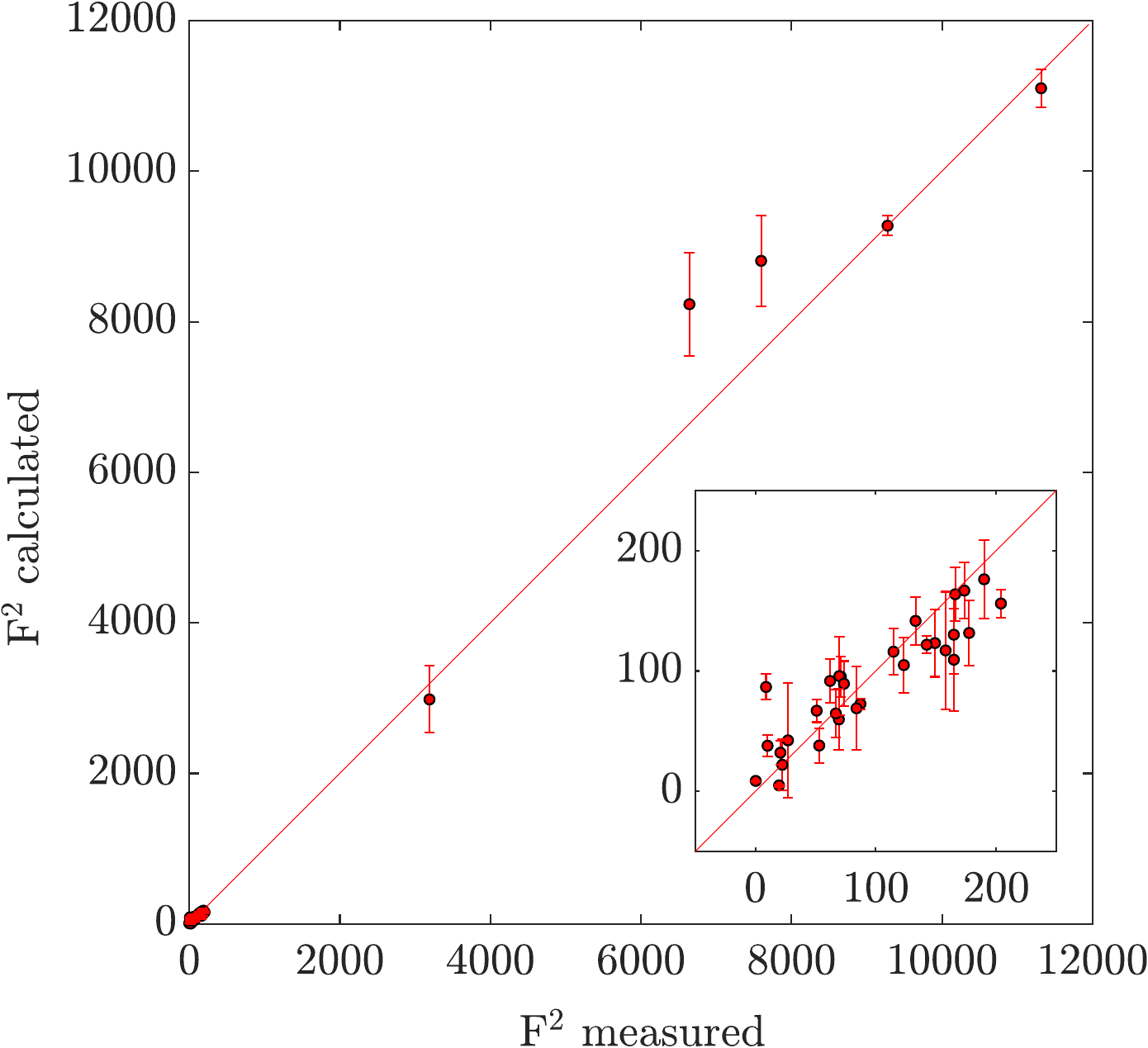}
\caption{Magnetic scattering measured on D10+. The inset shows the refinement of the weak reflections, associated mostly with the superstructure peaks.}\label{refinement_ave_mine}
\end{figure}

In order to refine the magnetic structure at the lowest measured temperature we subtracted data collected at 20\,K from those collected at 1.7\,K. Our refinement demonstrates that the measured intensities, shown in Fig.~\ref{refinement_ave_mine}, may be described by a magnetic unit cell containing two magnetic moments of distinct magnitudes, one ``short'' and one ``long'', both oriented ferromagnetically along $\mathbf{c}$. The temperature dependence of the superstructure peaks, shown in Fig.~\ref{T_dependence}, occurs because the ratio of the magnitudes of these two moments changes as the temperature is lowered below $T_\mathrm{C}$. The integer-index peaks depend on the sum of the two moments, whereas the non-integer peaks depend on the difference. Accordingly, the ratio is largest at $T\simeq 8$\,K where the intensity of the non-integer peaks is a maximum. At lower temperatures, the intensity of the non-integer peaks decreases as the moments become more similar in magnitude.

\begin{figure}
\centering
\hspace{-1em}
\subfigure[Domain 1]{%
\includegraphics[trim=800 100 700 100, clip,width=2.8cm]{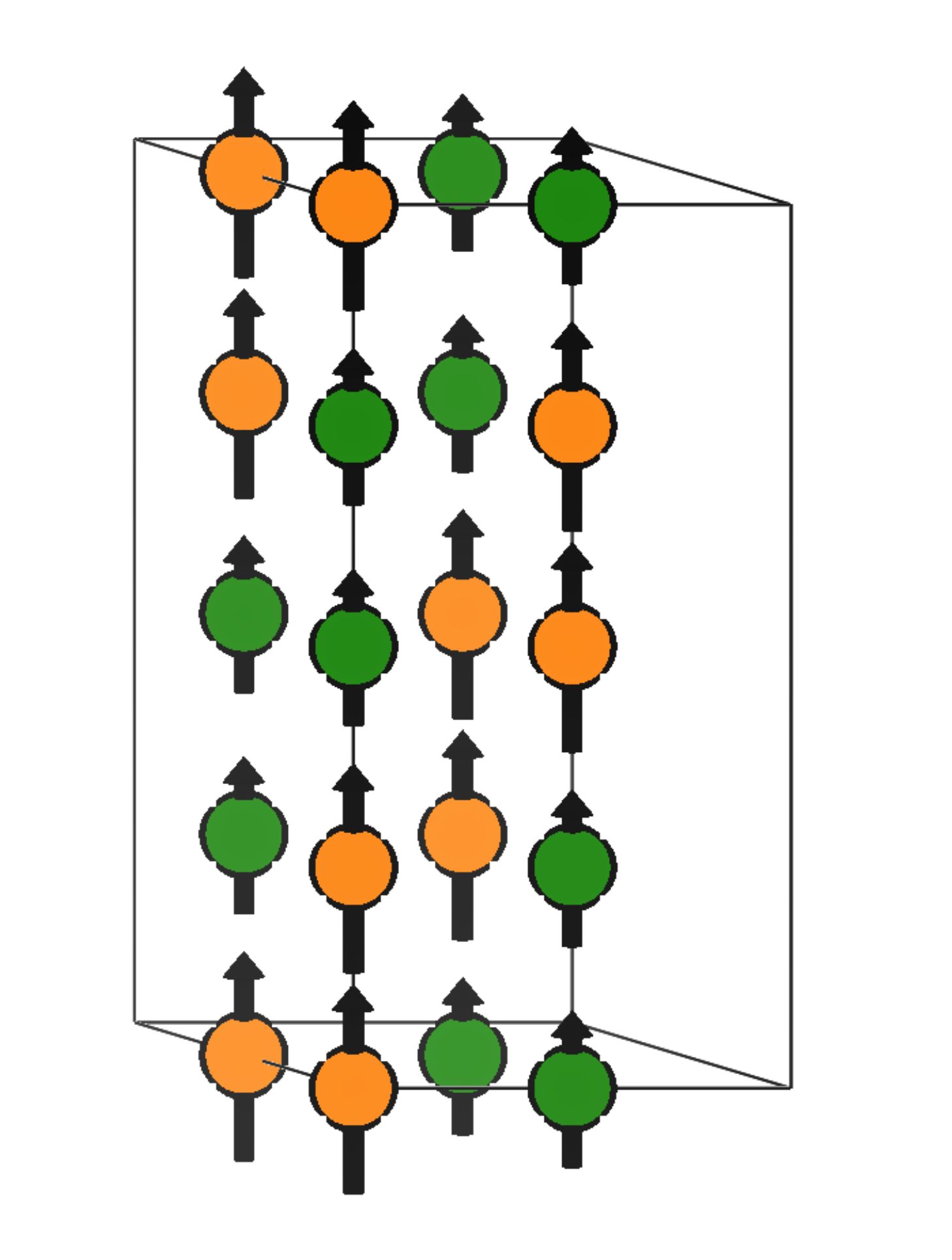}\label{dom1}}
\subfigure[Domain 2]{%
\includegraphics[trim=800 100 700 100, clip,width=2.8cm]{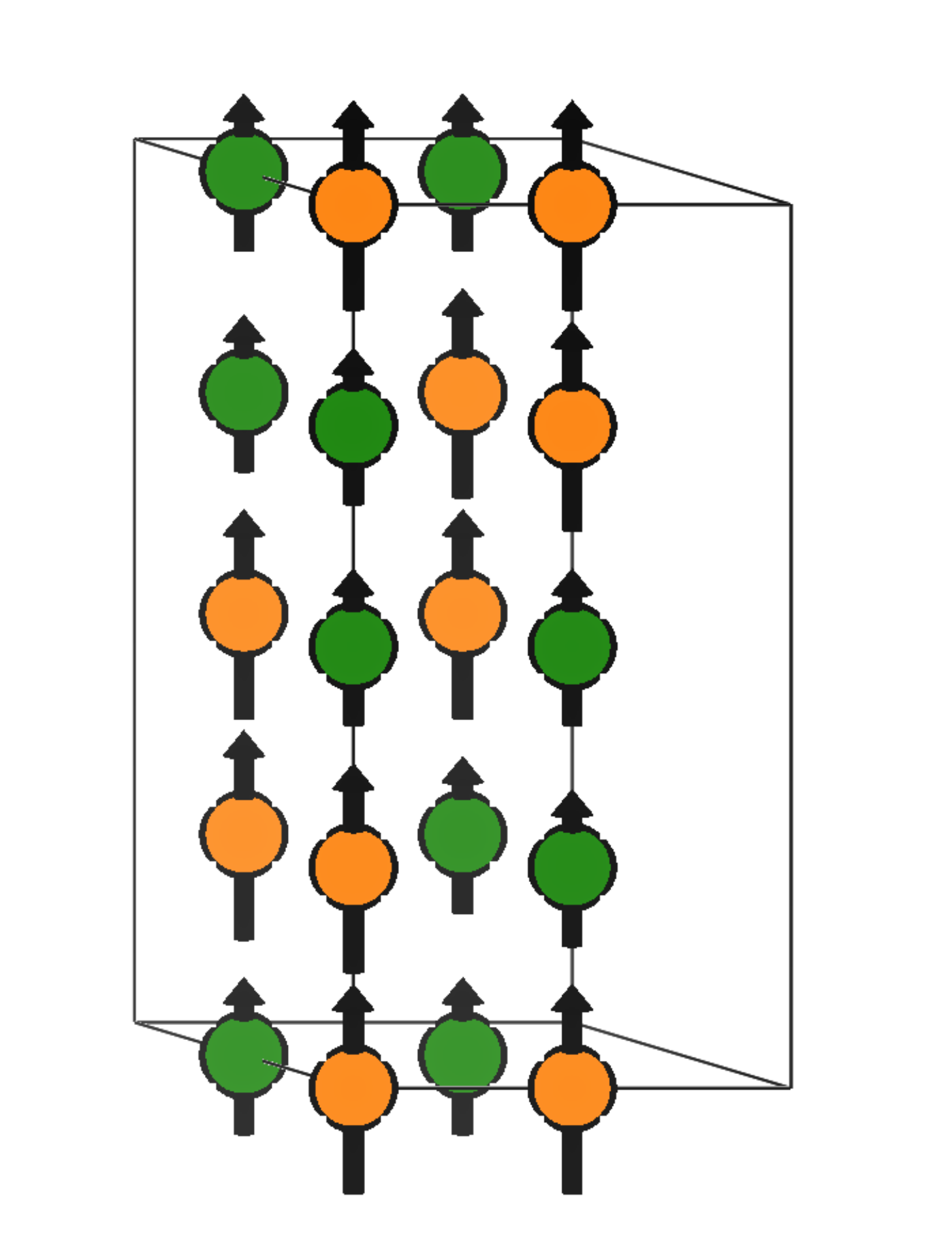}}
\subfigure[Domain 3]{%
\includegraphics[trim=800 100 700 100, clip,width=2.8cm]{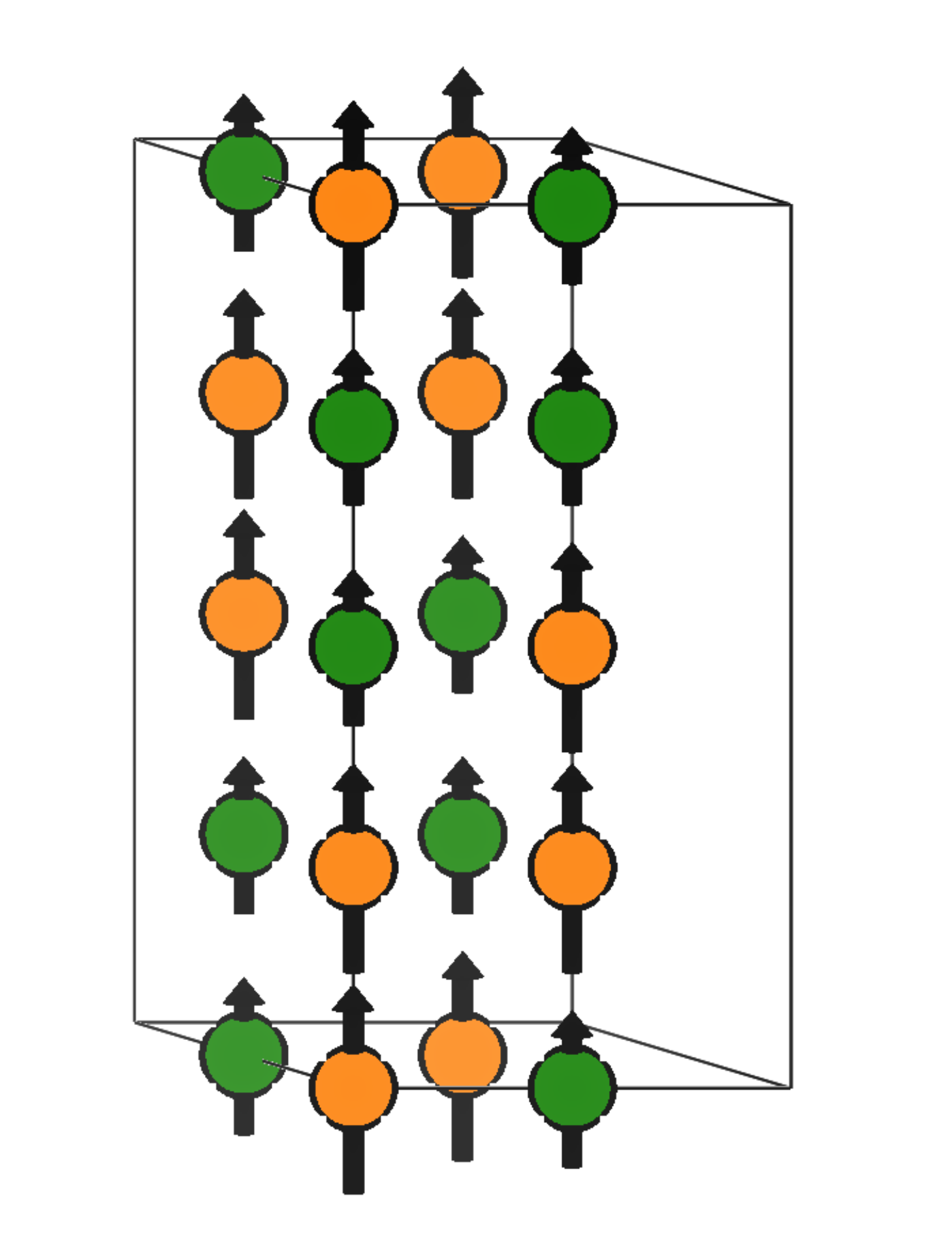}}
\subfigure[Domain 4]{%
\includegraphics[trim=800 100 700 100, clip,width=2.8cm]{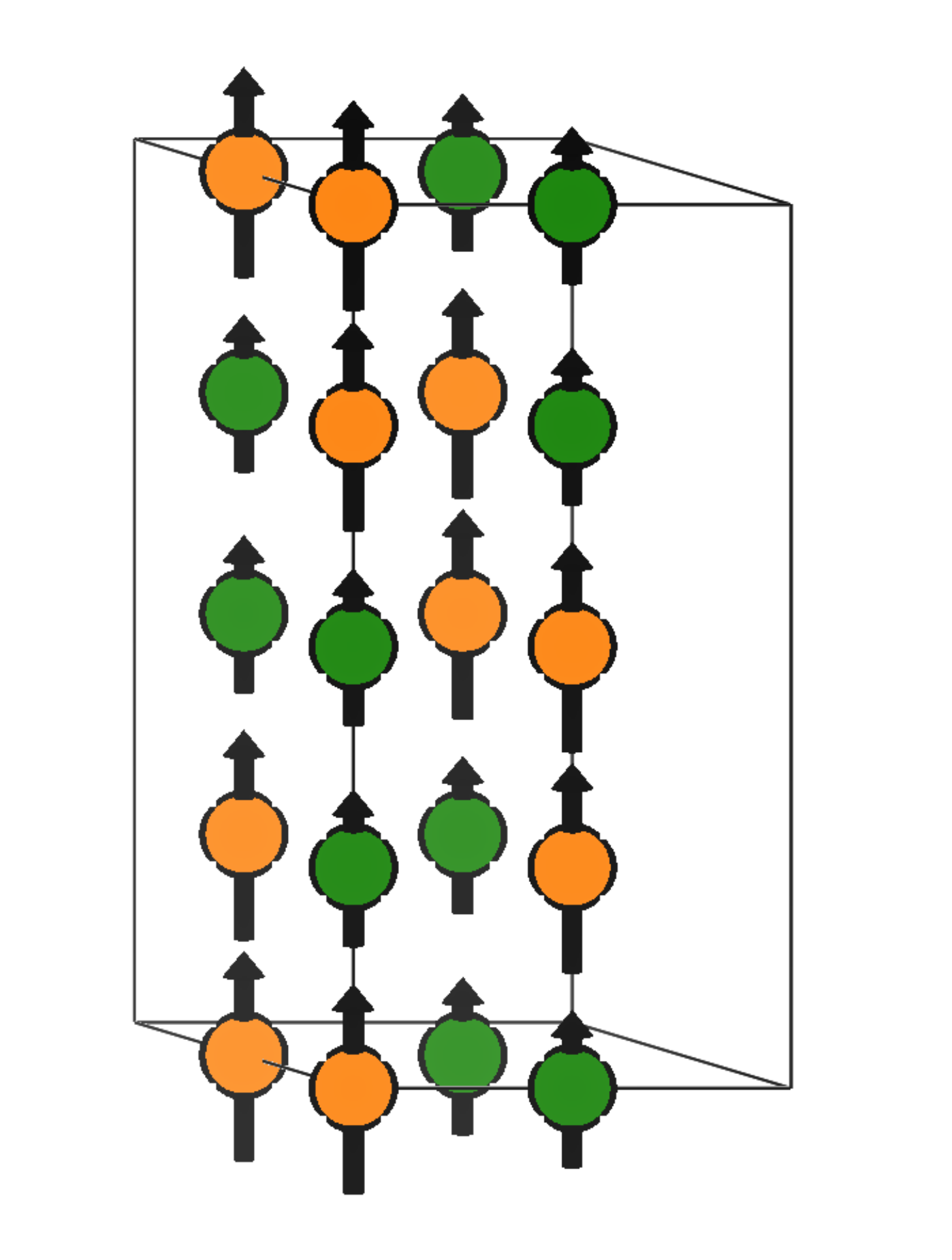}}
\subfigure[Domain 5]{%
\includegraphics[trim=800 100 700 100, clip,width=2.8cm]{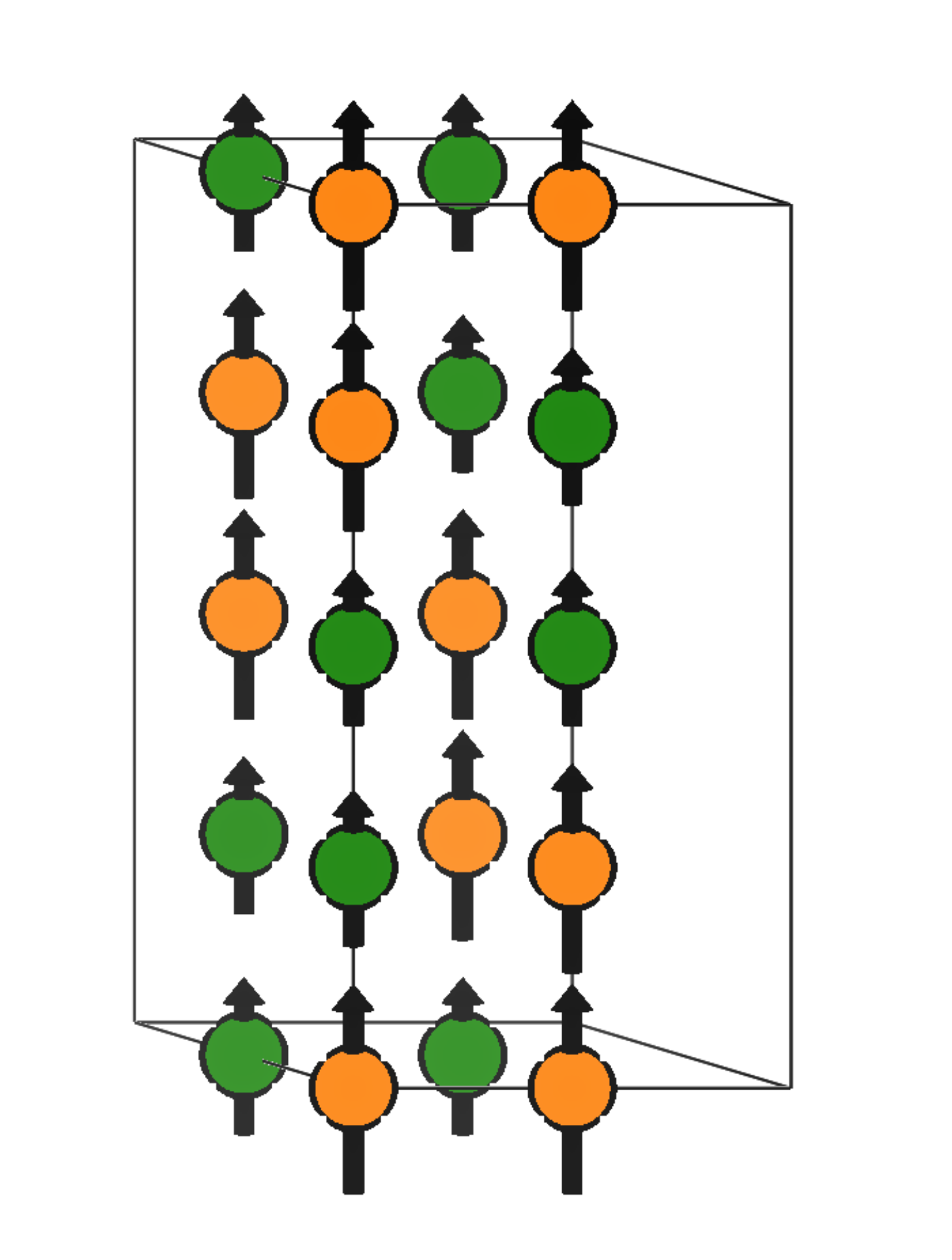}}
\subfigure[Domain 6]{%
\includegraphics[trim=800 100 700 100, clip,width=2.8cm]{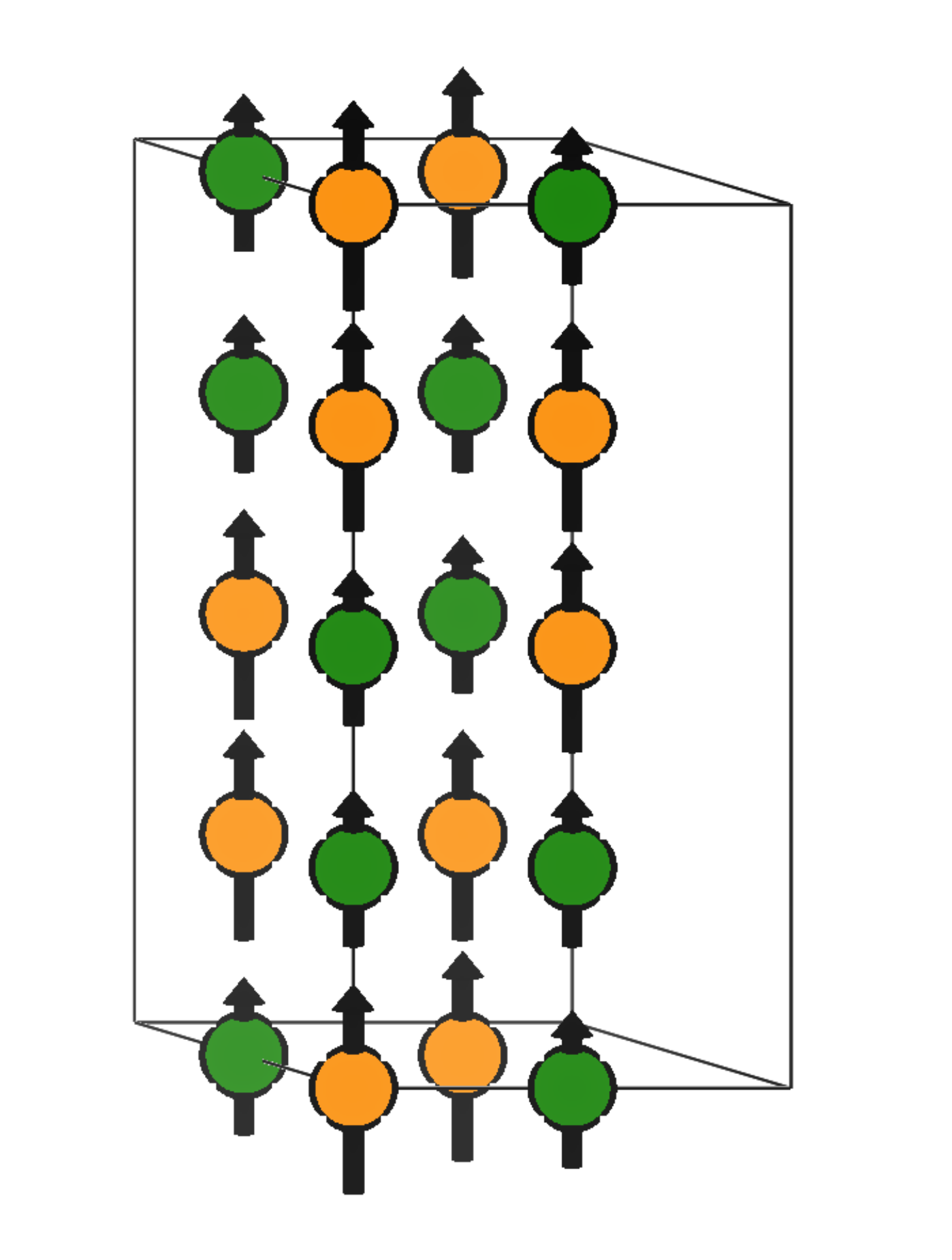}}
\caption{(a)-(f) Magnetic structure of the six domains of Nd$_2$PdSi$_3$. In our unit-cell origin choice, orange atoms have long, while green have short magnetic moments.}\label{Twins_magnetic}
\end{figure}

The refined magnetic structure using our model is represented in Fig.~\ref{Twins_magnetic} for each domain. We note that, from the diffraction experiment alone, it is impossible to know if the magnetic moments of the four-fold Pd-coordinated atoms are long and those of the two-fold Pd-coordinated atoms are short, or \textit{vice-versa}. The long and short ordered moments at 1.7\,K are found to be $2.3(3)\mu_{\mathrm{B}}$ and $1.3(3)\mu_{\mathrm{B}}$, respectively. 

The magnetic structure proposed here differs from that of Ref.~\onlinecite{PhysRevB.100.134423}. First, we do not find evidence for an incommensurate magnetic propagation vector $\mathbf{k}$, observed in their powder sample. Second, in our model the two different values of the Nd ordered moment correlate directly with the two different crystal-field environments caused by the Pd/Si superstructure, which is not the case for the model of Ref.~\onlinecite{PhysRevB.100.134423} as it assumes no superstructure. In fact, both models give the same diffraction intensities after domain averaging, but our model has a more physical basis with respect to the full crystallographic structure determined here.

\section{Crystal field parameters}

The $B_{q}^k$ parameters in our crystal-field model (see Eq.~(2) of the main text) for the long- and short-spin Nd sites are listed in Table~\ref{tab3}. Figures~\hyperref[tab3]{1(e)-(f)} in the main text illustrate the  energy-level splitting of the $4f$ single-ion states of Nd produced by these two sets of parameters. 

\begin{table}
\begin{ruledtabular}
\begin{tabular}{ccc}
Parameter & Long spin (meV) & Short spin (meV)\\
\colrule
$B_0^2$& \hspace{7pt}15 & \hspace{7pt}33 \\
$B_0^4$& $-23$ & $-33$  \\
$B_0^6$& \hspace{7pt}28 &  \hspace{7pt}31  \\
$B_6^6$& \hspace{4pt}$-1$ & $-12$  \\
\end{tabular}
\end{ruledtabular}
\caption{Fitted crystal field parameters for the two Nd sites in Nd$_2$PdSi$_3$. Parameters are given in the Wybourne convention.}\label{tab3}
\end{table}

\begin{figure*}
\centering
\subfigure{\includegraphics[trim=0 0 0 0, clip,width=5.5cm]{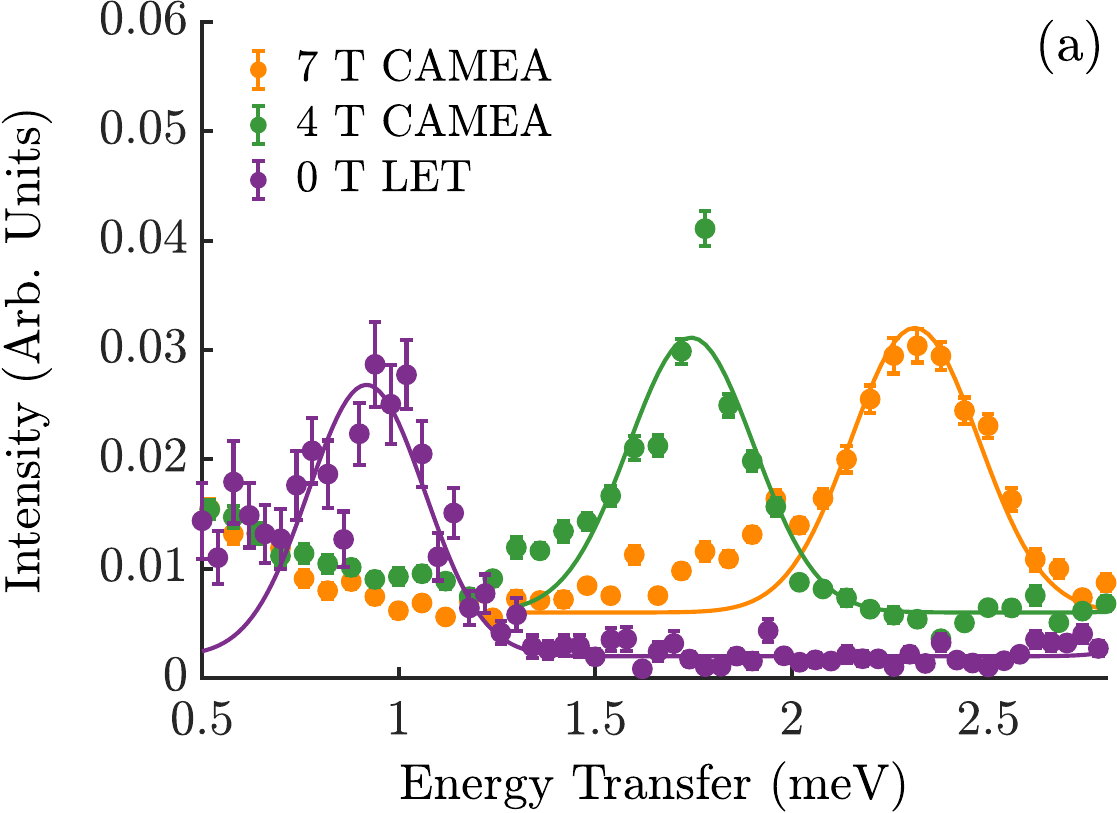}}
\subfigure{\includegraphics[trim=0 0 0 0, clip,width=5.5cm]{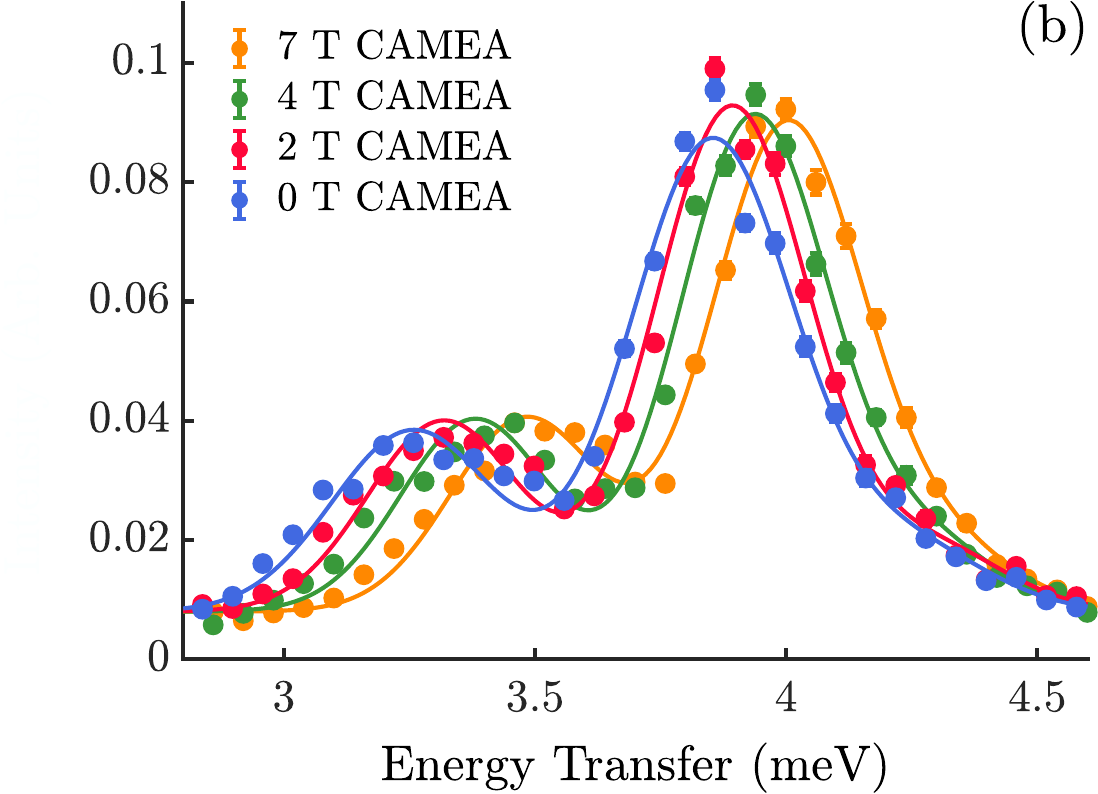}}
\subfigure{\includegraphics[trim=0 0 0 0, clip,width=5.5cm]{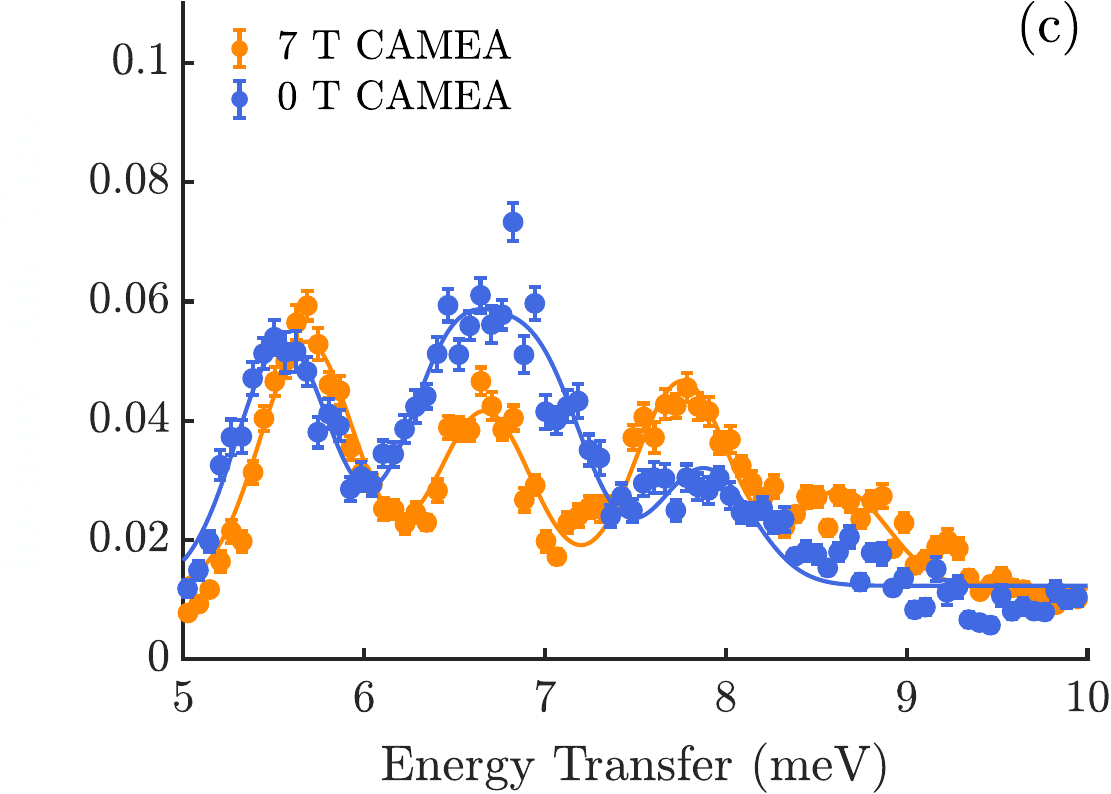}}
\caption{Constant-$\mathbf{Q}$ cuts performed at (a)-(b) $(-1/2, 1/2, 0)$ and (c) $(-1, 1, 0)$ r.l.u. on data collected on CAMEA at a temperature of $\sim 2$~K for several magnetic fields applied along $\mathbf{c}$. The data were integrated over $\pm0.15~\textrm {r.l.u.}$ along $\langle\mathbf{Q}\rangle$ and two perpendicular directions. Panel (a) includes the same cut from the LET data measured in zero field, for comparison. Solid lines are Gaussian fits which were used to identify the levels shift with field.}\label{camea}
\end{figure*}

Refinement of the CF parameters was informed by INS measurements of the response of the magnetic modes to an applied magnetic field. Fig.~\ref{camea} shows constant-\textbf{Q} cuts integrated around $(-1/2, 1/2, 0)$ and $(-1, 1, 0)$ r.l.u. for several different magnetic fields up to 11\,T applied parallel to $\mathbf{c}$. A shift of the CF levels towards higher energies occurs with increased field, and the rate of increase is different for each level. The largest energy shift observed is on the lowest mode, which results from the Zeeman splitting of the ground-state doublet of the short-spin site. The middle and bottom panels show the relatively weak field dependence of the three modes between 3--4\,meV, and the other levels measured above 5\,meV, respectively. In the latter energy range, the strongest observed field dependence is on the levels with energies above 6.8\,meV, which appear displaced by about 1\,meV for a 7\,T magnetic field. 

These data were included in our crystal-field analysis by the addition of a Zeeman splitting term
\begin{equation}
\mathcal{H}_\mathrm{z}=-\mu_\mathrm{B}g_{J}\mathbf{J}_{i}\cdot \mathbf{H},
\label{eq1}
\end{equation}
where $\mu_\mathrm{B}$ is the Bohr magneton, $g_{J}=8/11$ is the Land\'e $g$-factor and $\mathbf{H}$ is the magnetic field, to the Hamiltonian ~\hyperref[eq1]{Eq. (1)} in the main text. Gaussians were fitted to the whole dataset, including those shown in Fig.~\ref{camea}, and the measured splittings were used as fitting constraints to the CF model. 

\section{Exchange coupling}

\begin{figure*}
\centering
\includegraphics[trim=0 0 0 0, clip,width=12.2cm]{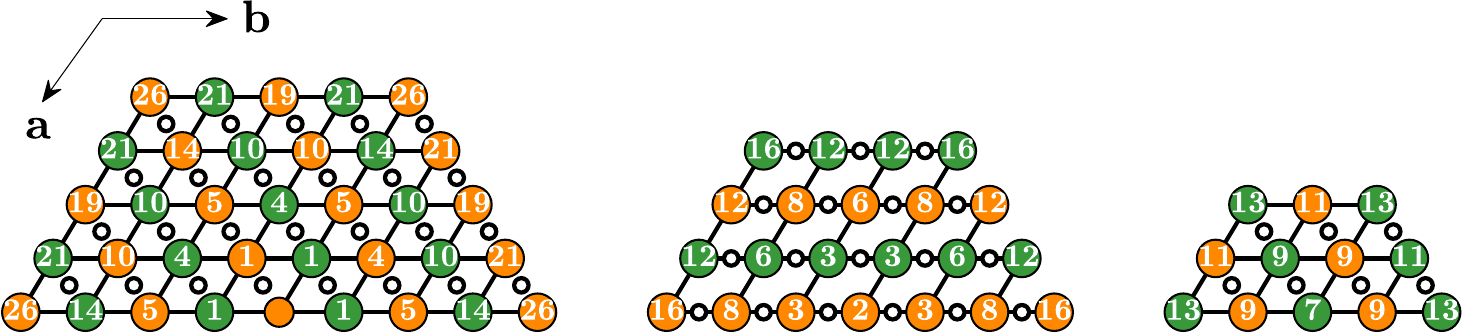}
\caption{Diagrams showing up to the 26$^\textrm{th}$ nearest-neighbor exchange pathways for superstructure domain 1 of Nd$_2$PdSi$_3$. The paths start on a long-spin Nd site (unnumbered, smaller orange circle in the diagram on the left). White circles represent inversion centres in the $\mathbf{ab}$-plane. The $c$ coordinate of each of the planes increases from left to right as one moves to the next layer.}\label{coupling}
\end{figure*}

Figure~\ref{coupling} shows, for domain 1 [Fig.~\hyperref[Twins_magnetic]{7(a)}], the position of each $n^\textrm{th}$-nearest neighbor of a long-spin Nd. In this work, couplings up to the 26$^\textrm{th}$ nearest neighbours were found to affect, to a lesser or greater extent, the observed dispersions of the magnetic levels of the two inequivalent rare-earth. Open white circles in Fig.~\ref{coupling} mark the position of inversion centres, which in the orthorhombic structure exist in the Nd layers between same-type sites.

Although all the evidence available points towards an orthorhombic description for Nd$_2$PdSi$_3$ as being the most accurate one, the data may not always be sensitive to the full symmetry of the $Fddd$ space group. In order to develop a model for exchange couplings, we initially assumed the highest possible symmetry (that imposed by $P6/mmm$), and subsequently lowered it, differentiating between the $ll$, $ls$ and $ss$ paths, where $l$ and $s$ refer to Nd sites with long and short moments. Since these assumptions have proven to be a minimal requirement of the model, all the exchange matrices are labelled $\boldsymbol{\mathcal{J}}_n^{ij}$, where $i,j$ represent either $l$ or $s$. 

\begin{figure}
\subfigure{\includegraphics[trim=0 0 0 0, clip,width=8.2cm]{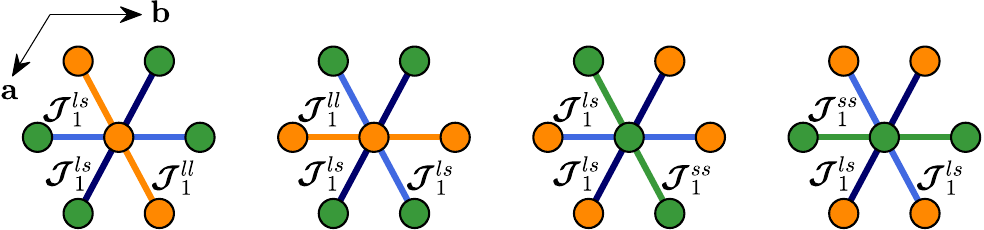}}
\subfigure{\includegraphics[trim=0 0 0 0, clip,width=8.2cm]{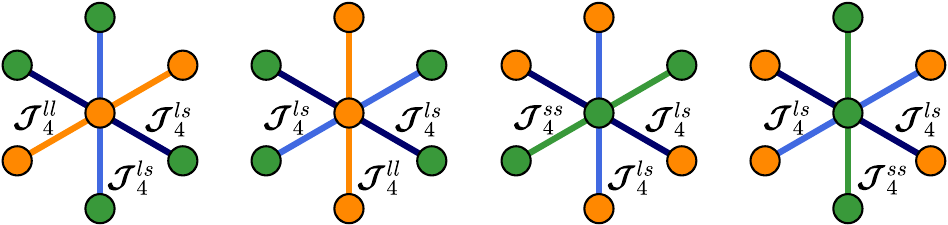}}
\subfigure{\includegraphics[trim=0 0 0 0, clip,width=8.2cm]{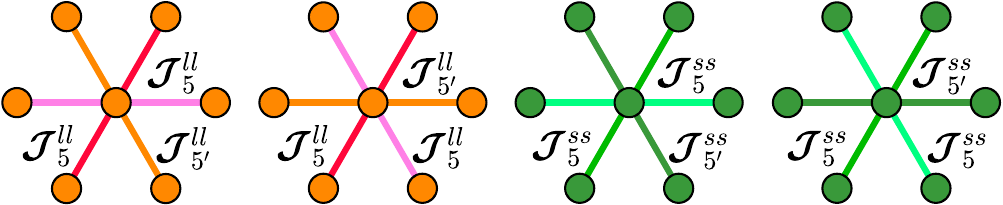}}
\caption{Diagrams showing the first-, fourth- and fifth- nearest-neighbor couplings between different Nd sites. Those correspond to first-, second- and third- nearest-neighbor bonds within the $ab$ plane, respectively. In these figures, the $c$ coordinate of each Nd increases from left to right. First central atom on the left side is located at $\tfrac{1}{2},\tfrac{1}{2},\tfrac{1}{4}$ in the conventional unit cell of Fig.~\ref{Layers}. Different colors show bonds which can be inequivalent in the $Fddd$ space group, whereas labels refer to the simplifying assumptions made in this work.}\label{layers_ab}
\end{figure}

\begin{figure}
\subfigure{\includegraphics[trim=0 0 0 0, clip,width=8.2cm]{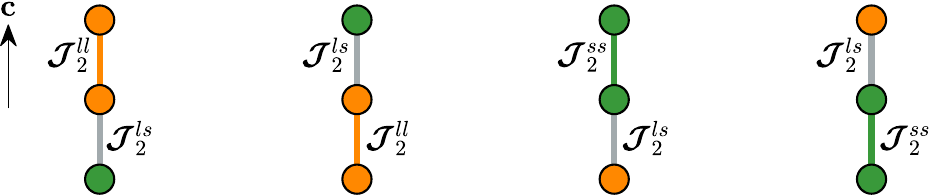}}
\subfigure{\includegraphics[trim=0 0 0 0, clip,width=4.0cm]{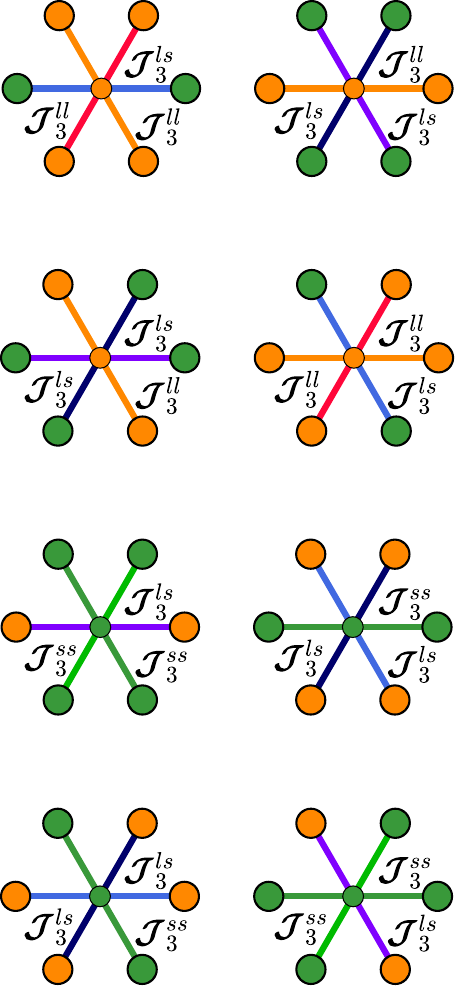}}
\caption{Diagrams showing the second- and third-nearest neighbor couplings between different Nd sites. As in Fig.~\ref{layers_ab}, different colors show bonds which can be inequivalent in the $Fddd$ space group. For $\boldsymbol{\mathcal{J}}_3$, left and right hexagons show layers below and above the central rare-earth, respectively.}\label{layers_c}
\end{figure}

Exchange pathways up to the $5^\textrm{th}$ nearest-neighbours are shown in Figs.~\ref{layers_ab} and \ref{layers_c}. The different colors of the bonds each correspond to a different $\boldsymbol{\mathcal{J}}_n^{ij}$ which would in principle be allowed in the $Fddd$ space group. For the $n=1,4$ in-plane couplings shown in Fig.~\ref{layers_ab}, for example, all the $\boldsymbol{\mathcal{J}}_{n}^{ll}$ and $\boldsymbol{\mathcal{J}}_{n}^{ss}$ are symmetry-related, but several different $\boldsymbol{\mathcal{J}}_{n}^{ls}$ interactions are allowed in one unit cell. When $n=3$, Fig.~\ref{layers_c}, there are different $\boldsymbol{\mathcal{J}}_n^{ij}$ for each of $ij = ll$, $ss$ and $ls$.

After careful investigation, we found that in most cases no improvement to the model could be achieved by allowing all the $\boldsymbol{\mathcal{J}}_n^{ij}$ for the same $n$ to vary independently. The only exceptions are for $n=5,19$ and 26, for which only ${ll}$ and ${ss}$ bonds exist, but not $ls$. As represented in Fig.~\ref{coupling}, some of the directions connecting the central atom with $n=5,19$ or $n=26$ neighbors contain inversion centres, while others do not. To mark the inequivalence of these two types of paths, bonds containing inversion are labelled as $\boldsymbol{\mathcal{J}}_{n'}^{ii}$, where $n'=5', 19 ', 26'$ and $i=l,s$. This notation was also used in Fig.~\ref{layers_ab}. All $\boldsymbol{\mathcal{J}}_n^{ls}$, $\boldsymbol{\mathcal{J}}_n^{ll}$ and $\boldsymbol{\mathcal{J}}_n^{ss}$ for $n\neq 5,19,26$ were considered to be the same apart from a similarity transformation (see next paragraph). In Figs.~\ref{layers_ab} and \ref{layers_c} the $\boldsymbol{\mathcal{J}}_n^{ij}$ labels indicate the bonds which we assumed to be the same.

If the coupling is of an isotropic Heisenberg type, which is our assumption when $n=2$ and $n \geq 5,5'$, an unitary transformation between any set of orthogonal coordinates will not modify the exchange matrices. The same is not true when the exchange is anisotropic. In this case, the coupling will depend on the basis and bond directions, although the trace of the matrices is always preserved. Therefore, before defining the exchange coupling matrices $\boldsymbol{\mathcal{J}}_n^{ij}$, we have to express the coordinate basis in which they are described. In this work's convention, the local orthonormal basis for a Nd atom located at $\big(1/2,1/2,1/4\big)$ in the $2\times2\times4$ conventional unit cell of domain 1 [Fig.~\ref{dom1}] is chosen to be $\hat{\mathbf{a}}_\mathrm{ex},\hat{\mathbf{b}}_\mathrm{ex},\hat{\mathbf{c}}_\mathrm{ex}$, where
\begin{align}
\begin{pmatrix}
\hat{\mathbf{a}}_\mathrm{ex} \\
\hat{\mathbf{b}}_\mathrm{ex} \\
\hat{\mathbf{c}}_\mathrm{ex} \\
\end{pmatrix}=
\renewcommand{\arraystretch}{1.3}
\begin{pmatrix}
1 \ & \ 0 \ & \ 0 \ \\
\frac{1}{\sqrt{3}} \ & \ \frac{2}{\sqrt{3}} \ & \ 0 \ \\
 0 \ & \ 0 \ & \ 1 \ \\
\end{pmatrix}
\begin{pmatrix}
\hat{\mathbf{a}} \\
\hat{\mathbf{b}} \\
\hat{\mathbf{c}} \\
\end{pmatrix},
\label{eq2}
\end{align}
and $\hat{\mathbf{a}},\hat{\mathbf{b}},\hat{\mathbf{c}}$ are unit vectors in the hexagonal basis, defined relative to the orthorhombic lattice in Fig.~\ref{Layers} above. This reference Nd (long-spin, orange) and its bonds are the first to be represented in Figs.~\ref{layers_ab} and \ref{layers_c}. For $\boldsymbol{\mathcal{J}}_{1}$ and $\boldsymbol{\mathcal{J}}_{3}$, the coordinate transformations are related to bond directions as follows:
\begin{itemize}
\item{$\boldsymbol{\mathcal{J}}_{n,b}^{ij}=3_{001}^\mathrm{T} \ \boldsymbol{\mathcal{J}}_{n,a}^{ij} 
\ 3_{001}$, where
\begin{align}
3_{001}=
\renewcommand{\arraystretch}{1.3}
\begin{pmatrix}
-\tfrac{1}{2} \ & \ \tfrac{\sqrt{3}}{2} \ & \ 0 \ \\
-\tfrac{\sqrt{3}}{2} \ & \ -\tfrac{1}{2} \ & \ 0 \ \\
 0 \ & \ 0 \ & \ 1 \ \\
\end{pmatrix}
\nonumber
\end{align}}
\end{itemize}
in the $(\hat{\mathbf{a}}_\mathrm{ex},\hat{\mathbf{b}}_\mathrm{ex},\hat{\mathbf{c}}_\mathrm{ex})$ coordinate basis and
\begin{itemize}
\item{$\boldsymbol{\mathcal{J}}_{n,ab}^{ij}=2_{100}^\mathrm{T} \ \boldsymbol{\mathcal{J}}_{n,b}^{ij} \ 2_{100}$, where
\begin{align}
2_{100}=
\begin{pmatrix}
1 \ & \ 0 \ & \ 0 \ \\
0 \ & \ -1 \ & \ 0 \ \\
0 \ & \ 0 \ & \ -1 \ \\
\end{pmatrix}
\nonumber
\end{align}}
\end{itemize}

Similarly, for $\boldsymbol{\mathcal{J}}_{4}$:
\begin{itemize}
\item{$\boldsymbol{\mathcal{J}}_{n,2ab}^{ij}=3_{001}^\mathrm{T} \ \boldsymbol{\mathcal{J}}_{i,a2b}^{nm} \ 3_{001}$ and}
\item{$\boldsymbol{\mathcal{J}}_{n,\bar{a}b}^{ij}=2_{100}^\mathrm{T} \ \boldsymbol{\mathcal{J}}_{n,2ab}^{ij} \
2_{100}$.}
\end{itemize}
Note that the three-fold rotation symmetry between bonds is not enforced by the $Fddd$ space group, but related to the assumptions made in this work (see above).  

Finally, the matrices from which the exchange couplings in the whole unit cell can be generated are stated below. The parameters are given in $\mu\mathrm{eV}$.

\begin{itemize}
\item{$n=1$}
\begin{align}
\boldsymbol{\mathcal{J}}_{1,a}^{ls}=
\begin{pmatrix}
-7 \ & \ 0 \ & \ 0 \ \\
 0 \ & \ 9.5 \ & \ 0 \ \\
 0 \ & \ 0 \ & \ 7.5 \ \\
\end{pmatrix}\nonumber,
\end{align}
\begin{align}
\boldsymbol{\mathcal{J}}_{1,b}^{ss}=
\begin{pmatrix}
10 \ &  \ 0 \ & \ 0 \ \\
 0 \ & \ 10 \ & \ 0 \ \\
 0 \ &  \ 0 \ & \ 10 \ \\
\end{pmatrix}\nonumber
\end{align}
and
\begin{align}
\boldsymbol{\mathcal{J}}_{1,b}^{ll}=
\begin{pmatrix}
5.2 \ & \ 0 \ & \ 0 \ \\
 0 \ & \ 15.6 \ & \ 0 \ \\
 0 \ & \ 0 \ & \ 10.4 \ \\
\end{pmatrix}.\nonumber
\end{align}
\end{itemize}

\begin{itemize}
\item{$n=3$}
\begin{align}
\boldsymbol{\mathcal{J}}_{3,a}^{ss}=
\begin{pmatrix}
4.7 \ & \ 0 \ & \ 0 \ \\
0 \ & \ 2.9 \ & \ 0 \ \\
0 \ & \ 0 \ & \ 3.8 \ \\
\end{pmatrix},\nonumber
\end{align}
\begin{align}
\boldsymbol{\mathcal{J}}_{3,a}^{ls}=
\begin{pmatrix}
5.2 \ & \ 0 \ & \ 0 \ \\
0 \ & \ 1.3 \ & \ 0 \ \\
 0 \ & \ 0 \ & \ 2.6 \ \\
\end{pmatrix}\nonumber
\end{align}
and
\begin{align}
\boldsymbol{\mathcal{J}}_{3,a}^{ls}=
\begin{pmatrix}
-3.3 \ & \ 0 \ & \ 0 \ \\
0 \ & \ 3.3 \ & \ 0 \ \\
 0 \ & \ 0 \ & \ 3.3 \ \\
\end{pmatrix}.\nonumber
\end{align}
\end{itemize}

\begin{itemize}
\item{$n=4$}
\begin{align}
\boldsymbol{\mathcal{J}}_{4,a2b}^{ls}=
\begin{pmatrix}
4.8 \ & \ 0 \ & \ 0 \ \\
0 \ & \ 1.8 \ & \ 0 \ \\
 0 \ & \ 0 \ & \ 3.6 \ \\
\end{pmatrix},\nonumber
\end{align}
\begin{align}
\boldsymbol{\mathcal{J}}_{4,a2b}^{ss}=
\begin{pmatrix}
-1.6 \ & \ 0 \ & \ 0 \ \\
0 \ & \ 2.4 \ & \ 0 \ \\
 0 \ & \ 0 \ & \ 2.4 \ \\
\end{pmatrix}\nonumber
\end{align}
and
\begin{align}
\boldsymbol{\mathcal{J}}_{4,a2b}^{ll}=
\begin{pmatrix}
-5.0 \ & \ 0 \ & \ 0 \ \\
0 \ & \ 4.0 \ & \ 0 \ \\
 0 \ & \ 0 \ & \ 3.0 \ \\
\end{pmatrix}.\nonumber
\end{align}
\vspace{0.01cm}
\end{itemize}
All other coupling matrices are assumed isotropic. The trace of each exchange matrix, which was used to generate Fig.~3 in the main text, is given in Table~\ref{tab4}. 

\begin{table}
\begin{ruledtabular}
\begin{tabular}{l c c c r}
Coupling & ss ($\mu$eV) & ls ($\mu$eV) &  ll ($\mu$eV) & $r_n$ (\AA)\\
\colrule
$\boldsymbol{\mathcal{J}}_1$ & 30.0 & 10.0 & 31.2 & 4.11\\
$\boldsymbol{\mathcal{J}}_2$ & 6.0 & 6.0 & $-30.0$ & 4.21\\
$\boldsymbol{\mathcal{J}}_3$ & 11.4 & 3.3 & 9.1 & 5.88\\
$\boldsymbol{\mathcal{J}}_4$ & 3.2 & 10.2 & 2.0 & 7.12\\
$\boldsymbol{\mathcal{J}}_{5'}$ & $-4.5$ & - & $-6.3$ & 8.22\\
$\boldsymbol{\mathcal{J}}_5$ & $-2.7$ & - & $-4.5$ & 8.22\\
$\boldsymbol{\mathcal{J}}_6$ & 3.6 & $-0.3$ & $-1.8$ & 8.27\\
$\boldsymbol{\mathcal{J}}_7$ & - &  $-3$ & - & 8.42 \\
$\boldsymbol{\mathcal{J}}_8$ & 1.8 & $-1.2$ & 1.2 & 9.23\\
$\boldsymbol{\mathcal{J}}_9$ & 2.4 &  $<|0.3|$ & $<|0.3|$ & 9.37\\
$\boldsymbol{\mathcal{J}}_{10}$ & $-3.6$ & $-1.3$ & $-2.3$ & 10.87\\
$\boldsymbol{\mathcal{J}}_{14}$ & $-3.6$ & $-3.6$ & $-1.8$ & 12.33\\
$\boldsymbol{\mathcal{J}}_{19'}$ & $-3.6$ & - & $-1.8$ & 14.23\\
$\boldsymbol{\mathcal{J}}_{19}$ & $-3.6$ & - & $-1.8$ & 14.23\\
$\boldsymbol{\mathcal{J}}_{21}$ &  $<|0.3|$ & $ -0.9$ & $ -0.9$ & 14.82 \\
$\boldsymbol{\mathcal{J}}_{26'}$ & $ -0.9$ & - & $ -0.9$ & 16.44 \\
$\boldsymbol{\mathcal{J}}_{26}$ & $<|0.3|$ & - & $ -0.4$ & 16.44 \\
\end{tabular}
\end{ruledtabular}
\caption{Trace of the exchange matrices in Nd$_2$PdSi$_3$. The minimum level of uncertainty in the parameters is $\pm 0.3~{\mu\mathrm{eV}}$.}\label{tab4}
\end{table}

The presence of inversion centres between Nd sites of the same type implies that DM coupling along these directions is forbidden. Antisymmetric exchange was, however, considered for all the other bonds where it is allowed, but the agreement between the model and the data did not improve. 

\section{Additional data and model quality considerations}

\begin{figure*}
\centering
\includegraphics[trim=10 80 10 90,clip,width=13.0cm]{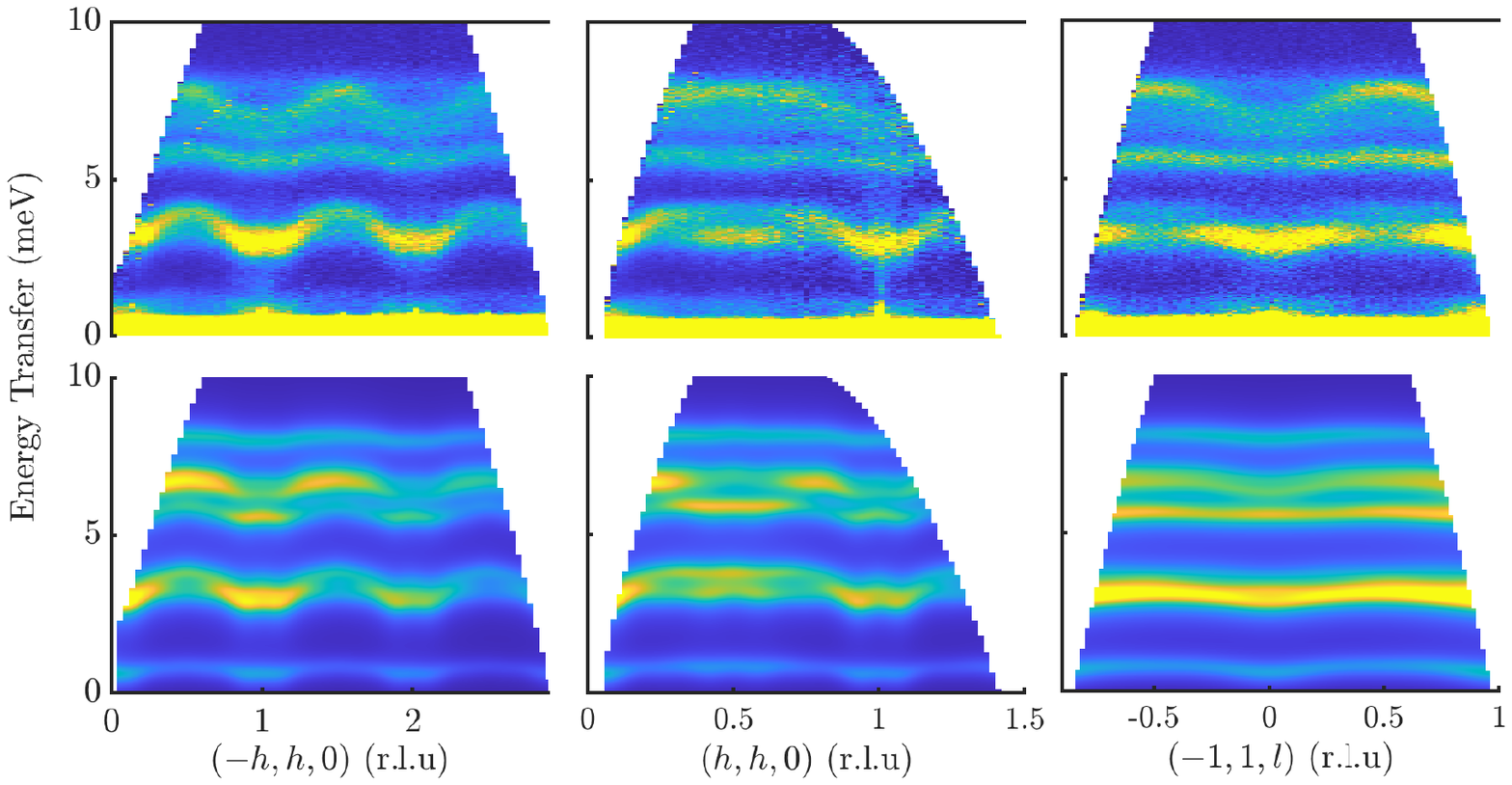}
\caption{Top panels show data collected on LET with neutrons of $E_\mathrm{i}=15.6~\mathrm{meV}$ displaying the coupled ground-$J$ multiplet of the Nd ion in Nd$_2$PdSi$_3$. Dispersions are seen propagating along three orthogonal directions: two in plane (left and centre, respectively) and one out-of-plane (right). Bottom panels show the coupled crystal-field and exchange model calculated using $\mathcal{H}$ defined in the main text.}\label{extra2}
\end{figure*}

Data measured on LET with neutrons of incident energy $E_\mathrm{i}=15.6~\mathrm{meV}$ along several high-symmetry directions in reciprocal space are shown in the upper panels of Fig.~\ref{extra2}. Similar data were collected on MERLIN with $E_\mathrm{i}=10~\mathrm{meV}$, in addition to chopper repetitions of $E_\mathrm{i}=20~\mathrm{meV}$ and $E_\mathrm{i}=50~\mathrm{meV}$ (not shown). The higher incident energies were used to ascertain that no excitations were present above $\sim 9~\mathrm{meV}$.

The calculated spectra of the exchange-coupled, ground-state $J$ multiplet of the Nd ion are shown along with the data in Fig.~\ref{extra2}. The energy of the levels up to $\sim5~\textrm{meV}$ is captured well by our reduced CF model, although the agreement is less good for the levels at higher energies. A better fit of these modes would necessarily require the use of an extended CF Hamiltonian, including terms allowed by symmetry but neglected in our calculation, as pointed out in the main text. This simplification of $\mathcal{H}_\mathrm{CF}$ means that in-plane single-ion anisotropy, which certainly exists in the real system -- see discussion below -- is not considered in the model. 

\begin{figure}
\centering
\includegraphics[trim=0 0 0 0, clip,width=6.5cm]{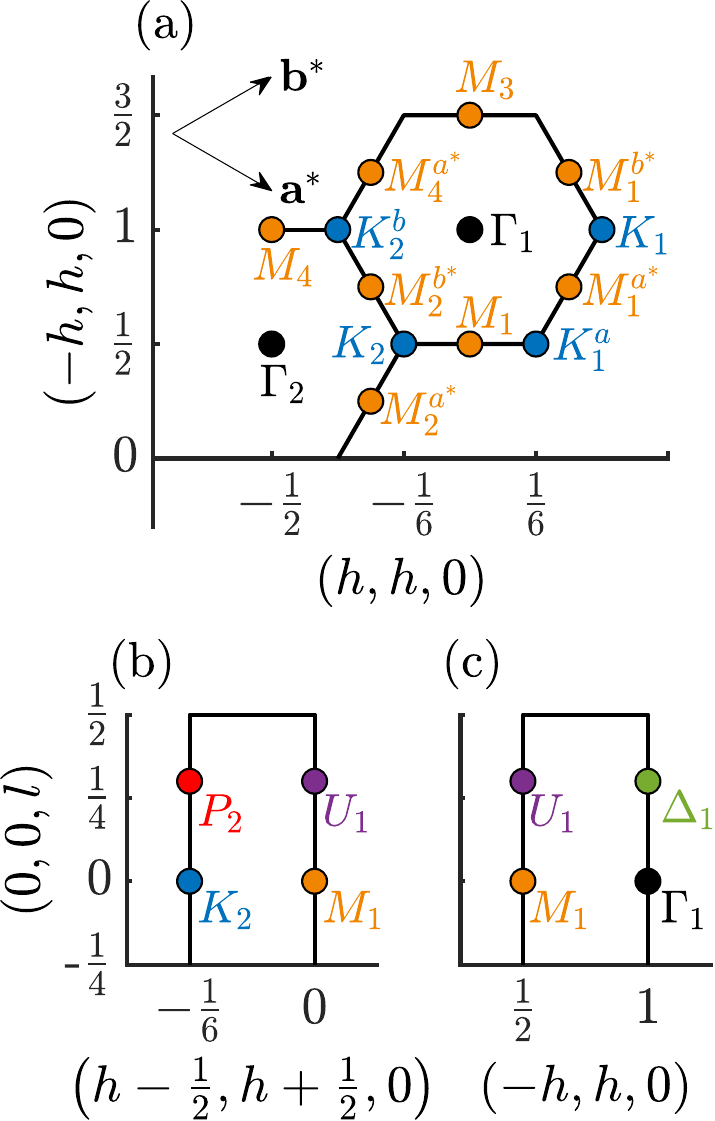}
\caption{Hexagonal Brillouin zone used in this work, showing labelled high-symmetry points which are inequivalent in the $Fddd$ space group. (a) shows in-plane and (b)--(c) show out-of-plane directions.}\label{BZs}
\end{figure}

Figure~\ref{BZs} shows the limits and definitions of the Brillouin zone used in the figure labels of the reciprocal space hypervolume measured with neutrons of $E_\mathrm{i}=6~\mathrm{meV}$. These refer to Fig.~2 in the main text and Fig.~\ref{extra1} below, which shows additional data measured away from $\Gamma$ points. Extra constant-energy cuts performed in addition to those shown in Fig.~4 of the main text can be seen in Fig.~\ref{aniso2}.

Figure~\ref{cq}, which contains constant-$\mathbf{Q}$ cuts, illustrates the level of agreement for several Brillouin zone points. The majority of the small details in the excitations are well described by our model, including the single-ion energies and dispersion of the modes. On the other hand, some simulations in Fig.~\ref{extra1} make evident the reciprocal space regions in which the calculation provides a less satisfactory description of the spectrum. One of the main differences between experiment and model is the intensity modulation caused by anisotropy in the energy level located between 3.6 and $3.8\,\mathrm{meV}$, seen in the constant-energy cuts of Fig.~\ref{aniso2}. Unlike what happened with the other levels appearing at similar energies, our attempts to fit the intensity of this mode considering only exchange anisotropy failed. A less good agreement is also seen in the out-of-plane directions, for which the data volume covered in the experiment is significantly smaller. The complete determination of exchange and single-ion anisotropies in the system is, in principle, possible, but would add an extra layer of complexity in an already very complicated problem, and would require additional data. 

\begin{figure*}
\includegraphics[trim=10 240 10 70,clip,width=17.5cm]{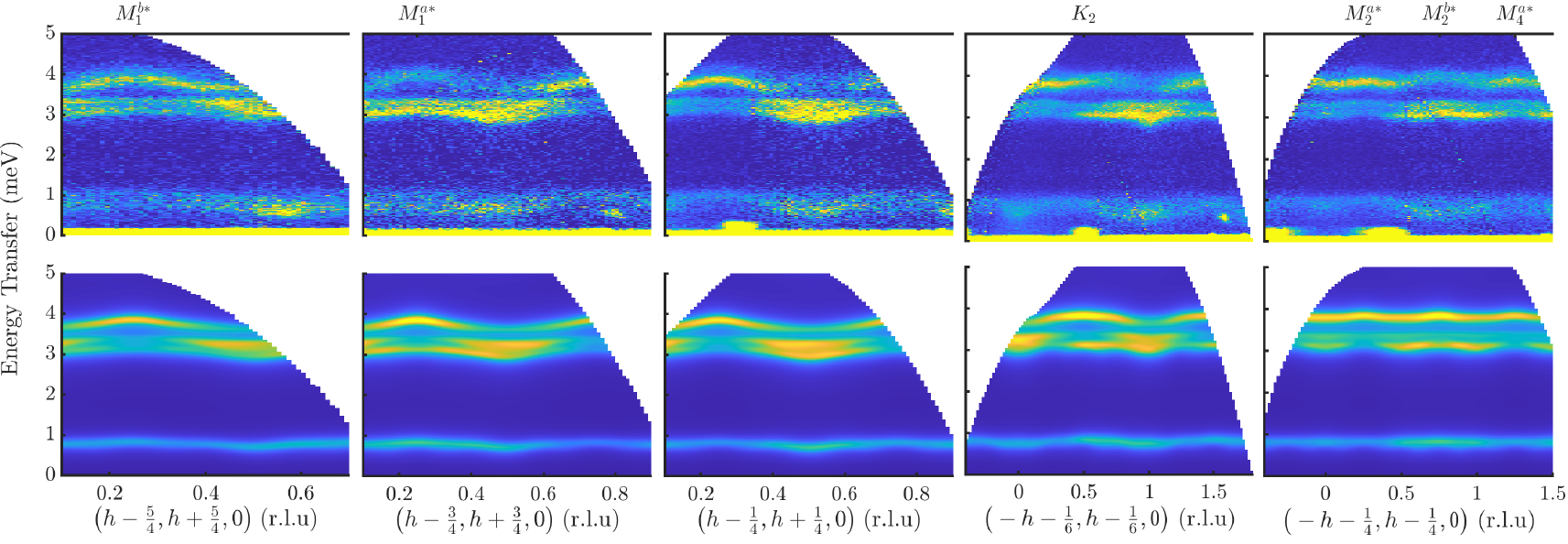}
\caption{Data collected on LET with neutrons of $E_\mathrm{i}=6~\mathrm{meV}$(top) along with calculated model (bottom) at a temperature of 1.7~K. Corresponding Brillouin zone labels are described in Fig.~\ref{BZs}.}\label{extra1}
\end{figure*}

\begin{figure}
\subfigure{\includegraphics[trim=5 30 28 0, clip,width=8.5cm]{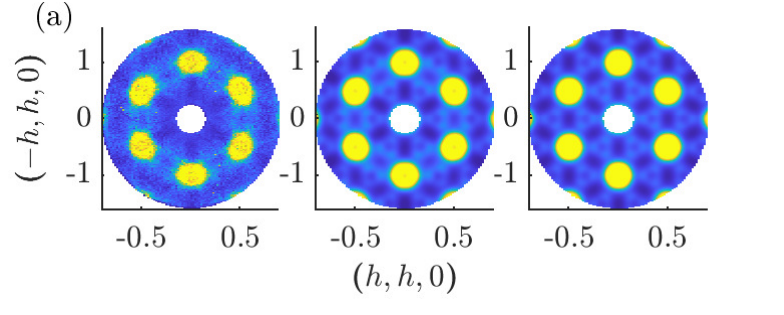}}
\subfigure{\includegraphics[trim=5 10 28 0, clip,width=8.5cm]{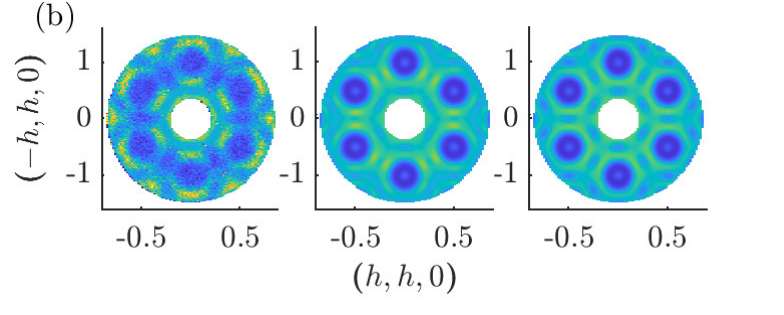}}
\caption{Constant-energy slices through data (left) and models (middle and right) integrated over the energy intervals (a) 2.7 to $3.0\,\mathrm{meV}$ and (b) 3.6 to $3.8\,\mathrm{meV}$, analogous to Fig. 4 in the main text. Middle and right panels show simulations using anisotropic and isotropic exchange coupling, respectively.}\label{aniso2}
\end{figure}

\begin{figure}
\centering
\includegraphics[trim=0 0 0 0, clip,width=8.3cm]{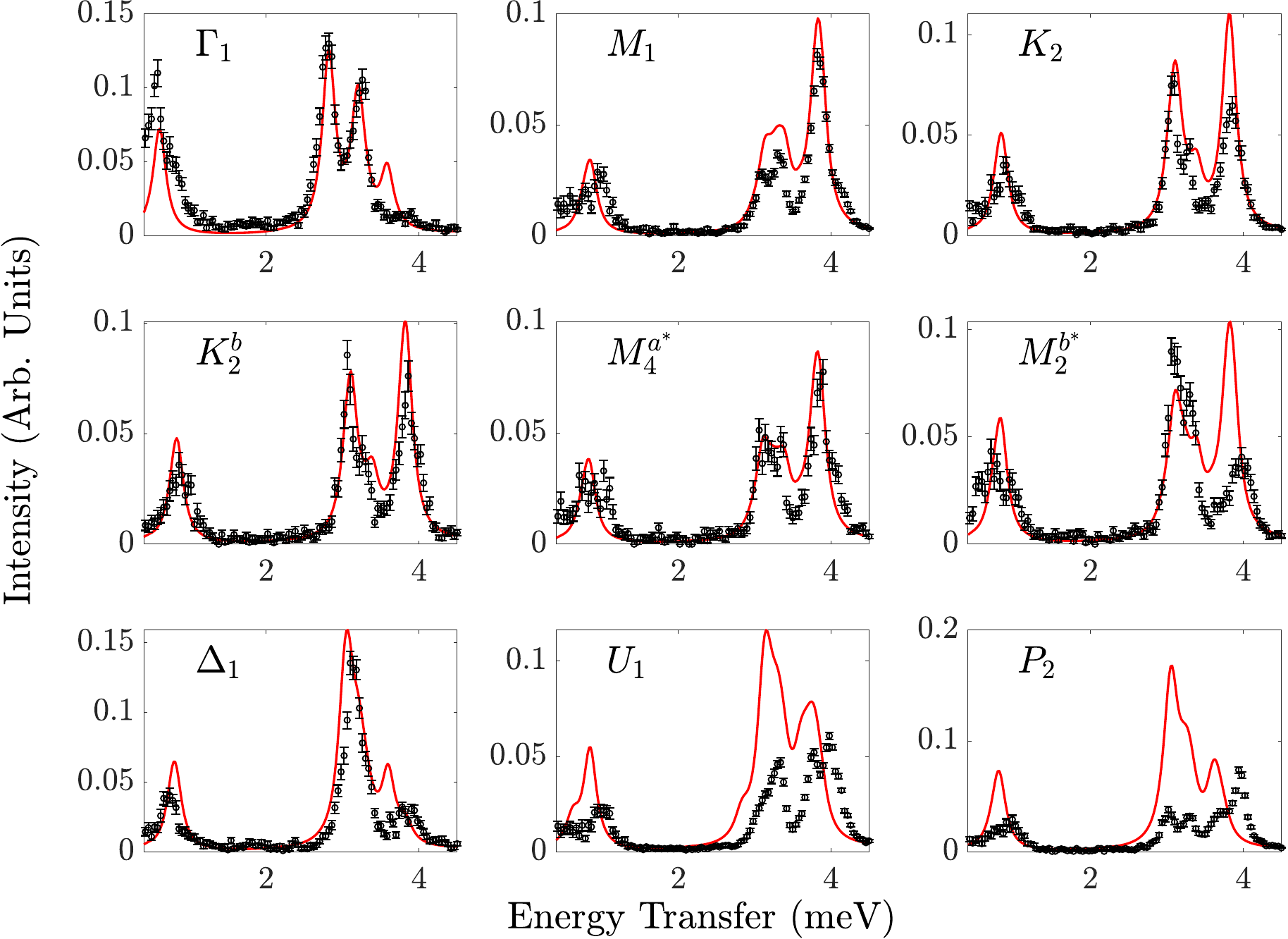}
\caption{Constant-$\mathbf{Q}$ cuts found by integrating scattering intensity measured at the reciprocal-lattice point written in the figure labels $\pm0.05$~r.l.u. in the three dimensional reciprocal space.}\label{cq}
\end{figure}

\section{Electronic structure calculations}

The electronic structure calculations for Nd$_2$PdSi$_3$ and Gd$_2$PdSi$_3$ were performed using a crystallographic unit cell with an orthorhombic space group \emph{Fddd} and a $2a\times2a\times4c$ superstructure. In other words, we assume that the superstructure found in Nd$_2$PdSi$_3$ applies for both compounds, to facilitate comparison of their electronic structures. Further information about the theoretical approach can be found in Refs. \onlinecite{PhysRevB.109.L201108,PhysRevLett.128.157206}. The lattice constants were taken from experiment and given in Table~\ref{Nd_Gd_comparison}, with five Wyckoff positions: one Pd, two rare-earth $R$, and two Si populated sites. The primitive cell contains 24 atoms with 4 Pd, 12 Si, and 8 rare-earth atoms. These rare-earth atoms are divided into two groups of four atoms each, $R_1$ and $R_2$. The electronic structure calculations are performed using multiple scattering (KKR) Density Functional Theory. The single-site potentials are generated using the HUTSEPOT DFT-KKR code \cite{pericles_15213951257}, which incorporates the local self-interaction correction on the 4$f$ electron orbitals \cite{PhysRevB.71.205109}. The 4$f$ electronic configuration is initially set to the Hund’s Rule state configuration, with a spin of $S = 3/2$ and orbital angular momentum $L = 6$ for Nd, and $S = 7/2$ and $L = 0$ for Gd. 

\begin{figure}
\centering
\includegraphics[trim=0 0 0 0, clip,width=8.4cm]{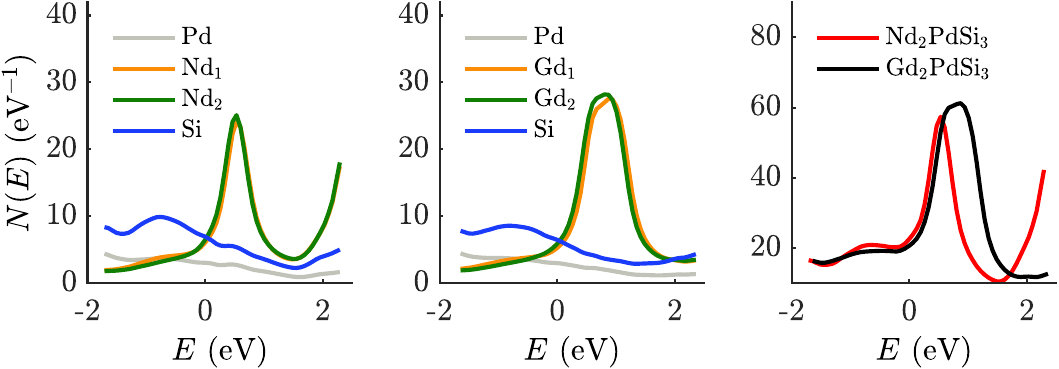}
\caption{Calculated element-specific density of states close to the Fermi energy ($E=0$\,eV) for Nd$_2$PdSi$_3$ (left) and Gd$_2$PdSi$_3$ (centre). Right: total density of states. The Nd$_2$PdSi$_3$ superstructure is assumed for both compounds.}\label{fig5}
\end{figure}

The local density of states (LDOS) was calculated, including relativistic effects, using the MARMOT electronic structure code \cite{Patrick_2022}. The DFT-KKR Green function is expanded using spherical harmonics with an orbital angular momentum cutoff $l_\textrm{max} = 3$. The LDOS is obtained using a mesh of energies with spacing of 0.005\,Ry and with an adaptive Brillouin zone integration with relative error over every division in wavevector space of less then 0.1\%. The LDOS near the Fermi energy is depicted in Fig.~5 in the main text and in Fig.~\ref{fig5} here. The latter shows that $R_1$ and $R_2$ are inequivalent (though very similar) for both compounds. This inequivalence stems from the different immediate surrounding environments [see main text, Fig.~1(a)], with $R_1$ surrounded by a hexagonal Si lattice in one adjacent layer, and the other hexagonal layer containing four Si atoms and two Pd atoms. In contrast, $R_2$ is surrounded by four Si atoms and two Pd atoms from both adjacent layers, leading to a different crystal-field environment. 

For Nd$_2$PdSi$_3$, the spin-up channel is split into two peaks. One of them is partially occupied, while the other is located $\sim0.5$\,eV above the Fermi energy (see Fig.~\ref{fig5}). For Gd$_2$PdSi$_3$, the $4f$ spin-up channel is fully occupied and more than 2\,eV below the Fermi energy, and the unoccupied $4f$ spin-down channel forms a single peak located around 1\,eV above the Fermi energy. The corresponding spin and orbital local moments obtained from the relativistic calculations are given in Table~\ref{Nd_Gd_comparison} for Nd$_2$PdSi$_3$ and Gd$_2$PdSi$_3$, respectively. The Nd compound has spin (orbital) moments close to 3\,$\mu_\text{B}$ (6\,$\mu_\text{B}$) while the Gd-one has large Gd spin moments (7\,$\mu_\text{B}$) as expected from Hund’s Rules. The tiny orbital moments on the Gd atoms originate from the spin polarization of the valence electrons (trivalent $5d^1$ $6s^2$). 

\begin{table*}
\centering
\begin{tabular}{lcc}
\hline
\textbf{Quantity} & \textbf{Nd$_2$PdSi$_3$} & \textbf{Gd$_2$PdSi$_3$} \\
\hline
Lattice Constants ($a, b, c$) (\AA) & (14.2345, 8.2183, 16.8442) & (14.0566, 8.1156, 16.3507) \\
\hline
Spin Moment ($\mu_B$) & (Nd$_1$: 3.2244, Nd$_2$: 3.2389) & (Gd$_1$: 7.0723, Gd$_2$: 7.0852) \\
\hline
Orbital Moment ($\mu_B$) & (Nd$_1$: -5.7582, Nd$_2$: -5.7743) & (Gd$_1$: 0.0747, Gd2: 0.0763) \\
\hline
\end{tabular}
\caption{Comparison of lattice constants and calculated spin and orbital moments for Nd$_2$PdSi$_3$ and Gd$_2$PdSi$_3$.}
\label{Nd_Gd_comparison}
\label{Nd_Gd_comparison}
\end{table*}

The $5d$ states derived from the rare-earth ions are very similar near the Fermi level for the Nd and Gd compounds, and the Si $3p$ and Pd $4d$ states near the Fermi energy are also very similar for the two compounds (see Fig.~5 in the main text and Fig.~\ref{fig5} here). The densities of states of the $6s$ and $6p$ bands from the rare earths are negligible compared with the other bands at the Fermi surface.

\bibliographystyle{apsrev4-2}
\bibliography{./bibs/PhysRevB.84.104105, 
./bibs/PhysRevB.100.134423,
./bibs/Kurumaji_science,
./bibs/Jens_Mackintosh_book,
./bibs/PhysRevLett.108.017206,
./bibs/PhysRevLett.124.207201,
./bibs/Leonov2015,
./bibs/Muhlbauer,
./bibs/Yu,
./bibs/Fert2013,
./bibs/Bogdanov2020,
./bibs/PhysRevLett.102.197202,
./bibs/Wiesendanger,
./bibs/PhysRevLett.127.067201,
./bibs/Moreau-Luchaire,
./bibs/Thomas,
./bibs/PhysRevX.9.041063,
./bibs/PhysRevLett.130.106701,
./bibs/Mallik_1998,
./bibs/KOTSANIDIS1990199,
./bibs/MALLIK1998169,
./bibs/PhysRevB.62.425,
./bibs/Frontzek_2007,
./bibs/FRONTZEK2006398,
./bibs/PhysRevB.68.012413,
./bibs/PhysRevB.62.14207,
./bibs/Nentwich_2016,
./bibs/SZYTULA1999365,
./bibs/PhysRevB.64.012418,
./bibs/PhysRevB.60.12162,
./bibs/PhysRevB.67.212401,
./bibs/Hirschberger_2019,
./bibs/Khanh_2020,
./bibs/Gao_2020,
./bibs/GORDON199724,
./bibs/CHEVALIER1984753,
./bibs/GLADYSHEVSKII1992221,
./bibs/PhysRevB.103.104408,
./bibs/PhysRevLett.125.117204,
./bibs/PhysRevB.95.224424,
./bibs/PhysRevB.93.064430,
./bibs/PhysRevLett.129.137202,
./bibs/MAYOH2024127774,
./bibs/LET,
./bibs/CARVAJAL1,
./bibs/CAMEA,
./bibs/MJOLNIR,
./bibs/Andrews_book,
./bibs/MERLIN,
./bibs/PhysRevB.109.L201108,
./bibs/PhysRevLett.128.157206,
./bibs/pericles_15213951257,
./bibs/PhysRevB.71.205109,
./bibs/Patrick_2022}